\documentclass[prb,tighten]{revtex4-1}

\usepackage{amssymb,amsmath}

\usepackage{color}

\usepackage{graphicx}
\newcommand{\dd}{\text{d}}

\begin{document}
\title{Grad's moment method for a granular fluid at moderate densities. Navier-Stokes transport coefficients}
\author{Vicente Garz\'{o}}
\email{vicenteg@unex.es}
\homepage{URL: http://www.unex.es/eweb/fisteor/vicente/}
\affiliation{Departamento de F\'{\i}sica, Universidad de Extremadura, E-06071 Badajoz, Spain}

\begin{abstract}
The Navier-Stokes transport coefficients of a granular dense fluid of smooth inelastic hard disks or spheres are explicitly determined by solving the inelastic Enskog equation by means of Grad's moment method. The
transport coefficients are explicitly determined as functions of the (constant) coefficient of restitution and the solid volume fraction. In addition, the cooling rate is also calculated to first order in the spatial gradients. The calculations are performed for an arbitrary number of dimensions. The results are not limited to small dissipation and are expected to apply at moderate densities. It is found that the expressions of the Navier-Stokes transport coefficients and the cooling rate agree with those previously obtained from the Chapman-Enskog method by using the leading terms in a Sonine polynomial expansion. This shows the equivalence between both methods for granular fluids in the Navier-Stokes approximation. A comparison with previous results derived from Grad's moment method for inelastic disks and spheres is also carried out.
\end{abstract}

\date{\today}
\maketitle

\section{Introduction}
\label{sec1}

Granular media under rapid flow conditions admit a hydrodynamic description.\cite{G03} The essential difference from that for ordinary fluids is the absence of energy conservation, yielding subtle modifications of the conventional Navier-Stokes (NS) equations for states with small spatial gradients of the hydrodynamic fields. To gain some insight into the general description of the dynamics of grains, a simple model is usually considered for a \emph{granular} gas: a system constituted by smooth hard spheres or disks with inelastic collisions. The loss of energy in each binary collision is accounted for by a \emph{constant} coefficient of normal restitution $\alpha \leq 1$, the case $\alpha=1$ corresponding to elastic collisions.

At a kinetic theory level, the Boltzmann \cite{BP04} and Enskog \cite{GS95,BDS97} equations conveniently adapted to account for inelastic binary collisions, have been employed in the past few years as the starting points to derive the NS hydrodynamic equations from a more fundamental point of view. In particular, assuming the existence of a \emph{normal} or hydrodynamic solution \cite{C90} for sufficiently long space and time scales, the Chapman-Enskog method \cite{CC70} has been applied to calculate the distribution function $f(\mathbf{r}, \mathbf{v}, t)$ through first order in the spatial gradients. Use of this distribution allows one to determine the dependence of the NS transport coefficients on the coefficient of restitution for dilute \cite{BDKS98} and moderately dense \cite{GD99,L05} gases. In contrast to previous attempts, \cite{LSJC84,L91,SG98} the results derived in Refs.\ \onlinecite{BDKS98,GD99,L05} do not impose any constraint on the degree of dissipation and take into account the (complete) nonlinear dependence of the transport coefficients on $\alpha$. However, as for elastic collisions, \cite{CC70} the exact forms of the NS transport coefficients require the solution of a set of linear integral equations and so the leading terms in a Sonine polynomial expansion (first Sonine approximation) are usually considered to get explicit expressions for all the above coefficients. In spite of this simple approach, the corresponding analytical results compare quite well with Monte Carlo simulations, \cite{MonteCarlo} except at high dissipation for the heat flux transport coefficients. \cite{BM04,MSG07} Motivated by this disagreement a modified version of the first Sonine approximation has been recently proposed \cite{GSM07,GVM09} that significantly improves the $\alpha$-dependence of the heat flux transport coefficients and corrects the discrepancies between simulation and theory.

An alternative procedure to solve the Boltzmann equation for a dilute gas is by means of the moment method. The objective of the method is to evaluate the velocity moments of the distribution function $f(\mathbf{r}, \mathbf{v}, t)$ rather than its explicit form as in the Chapman-Enskog method. Those velocity moments provide an indirect information on $f$ and, additionally, its first few moments (the mass density $\rho$, the flow velocity $\mathbf{U}$, the temperature $T$, the pressure tensor $P_{ij}$, and the heat flux vector $\mathbf{q}$) characterize the macroscopic state of the gas. Nevertheless, when one multiplies both sides of the Boltzman equation by a given set of functions $\psi (\mathbf{v})$ and integrates over the molecular velocity, in general one obtains an infinite \emph{hierarchy} of moment equations.\cite{C90} This infinite hierarchy can be recursively solved for some specific interactions potentials (such as Maxwell molecules, namely, when particles repel each other with a force inversely proportional to the fifth-power of the distance) and/or some special non-equilibrium situations. \cite{GS03} On the other hand, beyond this simple interaction potential, one has to resort to an approximate method to solve the above hierarchy of moment equations. The most widely used method was devised by Grad more than fifty years ago. \cite{G49} The idea of Grad's method is to assume $f(\mathbf{r}, \mathbf{v}, t)$ to be a local Maxwellian $f_\text{M}(\mathbf{r}, \mathbf{v}, t)$ times a sum over Hermite polynomials $H_k(\mathbf{v})$, i.e.,
\begin{equation}
\label{1.1}
f(\mathbf{r}, \mathbf{v}, t)\to f_\text{M}(\mathbf{r}, \mathbf{v}, t)\sum_{k=0}^{N-1}\; C_k(\mathbf{r},t)H_k(\mathbf{v}).
\end{equation}
The coefficients appearing in each of the velocity polynomials $H_k(\mathbf{v})$ are chosen by requiring that the corresponding velocity moments of Grad's solution \eqref{1.1} be the same as those of the exact velocity distribution function. There are $N$ arbitrary quantities which may be identified with the basic ($\rho$, $\mathbf{U}$, and $T$) and higher moments ($P_{ij}, \mathbf{q}, \cdots$) and can be determined by recursively solving the corresponding transfer equations for the above moments. A reasonable choice for a three-dimensional ordinary \emph{dilute} gas is $N=13$; in such a case the unknowns are the basic hydrodynamic fields ($\rho$, $\mathbf{U}$, and $T$) and the irreversible momentum and heat fluxes ($P_{ij}-nT \delta_{ij}$ and $\mathbf{q}$). In this case, the method is referred to as Grad's thirteen moment method. \cite{C90} In the case of a general dimensionality $d$ the number of involved moments is $d(d+5)/2+1$. Although Grad's moment method was originally proposed for dilute gases, its extension to dense gases is easy since one only has to consider the kinetic contributions to the fluxes in the trial solution \eqref{1.1}.

Grad's moment method has been also applied to granular gases. In the context of the \emph{inelastic} Enskog equation, Grad's thirteen-moment method was employed several years ago by Jenkins and Richman \cite{JR85a,JR85b} to determine the stress tensor, the heat flux, and the cooling rate in the NS approximation (linear theory). Although the application of Grad's method to the Enskog kinetic equation is not restricted to nearly elastic particles ($\alpha \simeq 1$), the results derived by Jenkins and Richman  \cite{JR85a,JR85b} neglect the cooling effects on the granular temperature $T$ due to the cooling rate. Given that this assumption can only be justified for nearly elastic systems, their expressions for the NS transport coefficients differ from those obtained in Refs.\ \onlinecite{GD99} and \onlinecite{L05} from the Chapman-Enskog method for \emph{arbitrary} degree of inelasticity. More recently, \cite{BST04} Grad's method has been also applied for weakly inelastic dilute gases with a coefficient of restitution which depends on the relative velocity. \cite{Nikolai}

The aim of this paper is to use Grad's method to obtain the NS transport coefficients of $d$-dimensional granular fluids described by the Enskog kinetic equation. This study extends a previous work of the author \cite{RGD12} for dilute granular gases and so it provides a description of hydrodynamics and transport at higher densities. With respect to the Jenkins-Richman theory, \cite{JR85a,JR85b} the present results incorporate two new ingredients not considered in the previous study. First, they are obtained by accounting for the time dependence of the temperature coming from the inelastic cooling. As a consequence, the corresponding expressions of the NS transport coefficients hold for arbitrary degree of dissipation. Second, a new scalar field (the full contracted moment of fourth order $c$) is added to the usual thirteen moments of mass density, velocity, temperature, and the kinetic contributions to the pressure tensor and heat flux vector. The inclusion of the fourth moment $c$ in a theory for ordinary gases that is related to Grad's moment method was proposed first by Kremer. \cite{K86} Subsequently, this moment has been also considered in some previous works \cite{RRSC00,K11} on dilute granular gases. As we will show later, a direct consequence of the presence of the field $c$ is a new contribution to the cooling rate proportional to the divergence of the flow velocity  as well as several new contributions to the transport coefficients coming from non-Gaussian corrections to the distribution function in the homogeneous cooling state.\cite{NE98} The question arises then as to wether, and if so to what extent, the conclusions drawn by Jenkins and Richman \cite{JR85a,JR85b} may be altered when the above two new ingredients are incorporated in Grad's solution.

In the NS approximation (namely, when only linear terms in the spatial gradients are retained), the constitutive equations for the pressure tensor $P_{ij}$, the heat flux ${\bf q}$ and the cooling rate $\zeta$ are \cite{GS95,GD99,JR85a,JR85b}
\begin{equation}
\label{1.2}
P_{ij}=p\delta_{ij}-\eta \left(\partial_{i}U_{j}+
\partial_{j}U_{i}-\frac{2}{d}\delta_{ij}\nabla \cdot \mathbf{U}\right)-\gamma \delta_{ij} \nabla \cdot \mathbf{U},
\end{equation}
\begin{equation}
\label{1.3}
\mathbf{q}=-\kappa \nabla T-\mu \nabla n,
\end{equation}
\begin{equation}
\label{1.4}
\zeta=\zeta_0+\zeta_\text{U} \nabla \cdot \mathbf{U},
\end{equation}
where $p$ is the hydrostatic pressure and $n$ is the number density. In addition, $\eta$ is the shear viscosity, $\gamma$ is the bulk viscosity, $\kappa$ is the thermal conductivity, $\mu$ is a new transport coefficient not present for elastic collisions, and $\zeta_0$ and $\zeta_\text{U}$ are the zeroth- and first-order contributions, respectively, to the cooling rate. The above NS transport coefficients can be written in reduced forms as
\begin{equation}
\label{1.5}
p(\alpha,\phi)=nT p^*(\alpha,\phi), \quad \eta(\alpha,\phi)=\eta(1,\phi) \eta^*(\alpha,\phi), \quad \gamma(\alpha,\phi)=\eta(1,\phi) \gamma^*(\alpha,\phi),
\end{equation}
\begin{equation}
\label{1.6}
\kappa(\alpha,\phi)=\kappa(1,\phi)\kappa^*(\alpha,\phi), \quad \mu(\alpha,\phi)=\frac{T\kappa(1,\phi)}{n}\mu^*(\alpha,\phi), \quad
\zeta_0(\alpha,\phi)=\frac{nT}{\eta(1,\phi)}\zeta_0^*(\alpha,\phi) ,
\end{equation}
where
\begin{equation}
\label{1.7}
\phi=
\frac{\pi^{d/2}}{2^{d-1}d\Gamma \left(\frac{d}{2}\right)}n\sigma^d,
\end{equation}
is the solid volume fraction and $\sigma$ is the diameter of the spheres. The coefficients $\eta(1,\phi)$ and $\kappa(1,\phi)$ are the values of the \emph{elastic} shear viscosity and thermal conductivity, respectively, given by the Enskog equation. \cite{FK72} The results derived in this paper show that the dimensionless coefficients $p^*$, $\eta^*$, $\gamma^*$, $\kappa^*$, $\mu^*$, $\zeta_0^*$, and $\zeta_\text{U}$ are in general \emph{nonlinear} functions of the coefficient of restitution $\alpha$ and the solid volume fraction $\phi$. In addition, the expressions of \emph{all} the above dimensionless NS transport coefficients obtained here from Grad's moment method agree with those derived from the Chapman-Enskog expansion in the first Sonine approximation. \cite{GD99,L05} This confirms the expected mutual consistency between both methods for solving the \emph{inelastic} Enskog equation in the NS domain.

The plan of the paper is as follows. In Sec.\ \ref{sec2} the Enskog kinetic equation and associated macroscopic conservation laws for a granular fluid are introduced. An overview of Grad's moment method used for solving this kinetic equation is given in Sec.\ \ref{sec3}. The explicit results for the NS transport coefficients are provided in Sec.\ \ref{sec4}, with the details of the calculations appearing in Appendices \ref{appA}, \ref{appB}, and \ref{appC}. A comparison with the results obtained by Jenkins and Richman for disks \cite{JR85a} and spheres \cite{JR85b} is done in Sec.\ \ref{sec5}, showing significant discrepancies between both theories especially in the heat flux transport coefficients. Finally, a short summary of the results derived in the paper is presented in Sec.\ \ref{sec6}.


\section{Enskog kinetic theory and conservation laws}
\label{sec2}

We consider a granular fluid composed by smooth inelastic disks or spheres of mass $m$ and diameter $\sigma$. The inelasticity of collisions among all pairs is accounted for by a \emph{constant} coefficient of normal restitution $0\leq \alpha \leq 1$ that only affects to the translational degrees of  freedom of grains. The particular value $\alpha=1$ corresponds to elastic collisions (ordinary fluids). At a kinetic theory level, all the relevant information on the state of the system is given by the one-particle velocity distribution function $f(\mathbf{r}, \mathbf{v}, t)$. For moderate densities, the inelastic Enskog theory \cite{BDS97} gives the time evolution of $f(\mathbf{r}, \mathbf{v}, t)$. In the absence of an external force, the Enskog equation has the form
\begin{equation}
\left(\partial_{t}+\mathbf{v}\cdot \mathbf{\nabla}\right)f(\mathbf{r},\mathbf{v},t)=
J_\text{E}\left[\mathbf{r},{\bf v}|f(t)\right], \label{2.1}
\end{equation}
where
\begin{equation}
\label{2.2}
J_{\text{E}}\left[{\bf r}, {\bf v}_{1}|f(t)\right] =\sigma^{d-1}\int \dd{\bf v}
_{2}\int \dd\widehat{\boldsymbol{\sigma}}\,\Theta (\widehat{{\boldsymbol {\sigma}}}
\cdot {\bf g})(\widehat{\boldsymbol {\sigma }}\cdot {\bf g})\left[ \alpha^{-2}
f^{(2)}({\bf r},{\bf r}-\boldsymbol {\sigma}, {\bf v}_1', {\bf v}_2';t)-
f^{(2)}({\bf r},{\bf r}+\boldsymbol {\sigma}, {\bf v}_1, {\bf v}_2;t)\right]
\end{equation}
is the Enskog collision operator. In Eq.\ \eqref{2.2}, $d$ is the dimensionality of
the system ($d=2$ for disks and $d=3$ for spheres), $\boldsymbol
{\sigma}=\sigma \widehat{\boldsymbol {\sigma}}$, $\widehat{\boldsymbol
{\sigma}}$ being a unit vector along the centers of the two colliding spheres,
$\Theta $ is the Heaviside step function, ${\bf g}={\bf v}_{1}-{\bf v}_{2}$ is the relative velocity and
\begin{equation}
\label{2.3}
f^{(2)}({\bf r}_1, {\bf r}_2, {\bf v}_1, {\bf v}_2, t)\equiv \chi({\bf r}_1, {\bf r}_2|n(t))f({\bf r}_1, {\bf v}_1, t) f({\bf r}_2, {\bf v}_2, t).
\end{equation}
The primes on the velocities in Eq.\ \eqref{2.2} denote the initial values $\{\mathbf{v}_1', \mathbf{v}_2'\}$ that lead to $\{\mathbf{v}_1, \mathbf{v}_2\}$ following a binary collision:
\begin{equation}
\label{2.3.1}
{\bf v}_{1}^{\prime}={\bf v}_{1}-\frac{1}{2}\left( 1+\alpha^{-1}\right)(\widehat{{\boldsymbol {\sigma }}}\cdot {\bf g})\widehat{{\boldsymbol {\sigma }}}, \quad
{\bf v}_{2}^{\prime }={\bf v}_{2}+\frac{1}{2}\left( 1+\alpha^{-1}\right)
(\widehat{{\boldsymbol {\sigma }}}\cdot {\bf g})\widehat{
\boldsymbol {\sigma}}.
\end{equation}
Furthermore, $\chi[{\bf r},{\bf r}+\boldsymbol{\sigma}|n(t)] $ is the equilibrium pair correlation function at contact as a functional of the non-equilibrium density field $n({\bf r}, t)$ defined by
\begin{equation}
\label{2.4}
n({\bf r}, t)=\int\; \dd{\bf v} f({\bf r},{\bf v},t).
\end{equation}

The first $d+2$ velocity moments of $f({\bf r},{\bf v},t)$ define the number density $n({\bf r}, t)$, the flow velocity
\begin{equation}
\label{2.5}
\mathbf{U}({\bf r}, t)=\frac{1}{n({\bf r}, t)}\int\; \dd{\bf v}\; \mathbf{v} f({\bf r},{\bf v},t),
\end{equation}
and the \emph{granular} temperature
\begin{equation}
\label{2.6}
T({\bf r}, t)=\frac{m}{d n({\bf r}, t)}\int\; \dd{\bf v}\; V^2 f({\bf r},{\bf v},t),
\end{equation}
where $\mathbf{V}({\bf r}, t) \equiv \mathbf{v}-\mathbf{U}({\bf r}, t)$ is the peculiar velocity.

The exact macroscopic balance equations for $n(\mathbf{r},t)$, $\mathbf{U}(\mathbf{r},t)$ and $T(\mathbf{r},t)$ follow directly from the Enskog equation \eqref{2.1} by multiplying with $1$, $m\mathbf{v}$, and $\frac{1}{2}mv^2$ and integrating over $\mathbf{v}$. After some algebra, one gets \cite{GD99}
\begin{equation}
\label{2.7}
D_t n+n\nabla \cdot \mathbf{U}=0,
\end{equation}
\begin{equation}
\label{2.8}
\rho D_t U_i+\partial_j P_{ij}=0,
\end{equation}
\begin{equation}
\label{2.9}
D_t T+\frac{2}{dn}\left(\partial_i q_i+ P_{ij}\partial_j U_i \right)=-\zeta T,
\end{equation}
where $D_t\equiv \partial_t+\mathbf{U}\cdot \nabla$ is the material derivative and $\rho=mn$ is the mass density. The
cooling rate $\zeta$ is (essentially) proportional to $1-\alpha^2$ and is due to
dissipative collisions. The pressure tensor ${\sf P}({\bf r},t)$ and
the heat flux ${\bf q}({\bf r},t)$ have both {\em kinetic} and {\em collisional transfer} contributions, i.e., ${\sf P}={\sf P}^k+{\sf P}^c$ and ${\bf q}={\bf q}^k+{\bf q}^c$. The kinetic contributions
are given by
\begin{equation}
\label{2.10}
{\sf P}^k({\bf r}, t)=\int \; \dd{\bf v} m{\bf V}{\bf V}f({\bf r},{\bf v},t), \quad
{\bf q}^k({\bf r}, t)=\int \; \dd{\bf v} \frac{m}{2}V^2{\bf V}f({\bf r},{\bf v},t),
\end{equation}
and the collisional transfer contributions are \cite{GD99}
\begin{equation}
{\sf P}^{c}({\bf r}, t)=\frac{1+\alpha}{4}m \sigma^{d}
\int \dd\mathbf{v}_{1}\int \dd\mathbf{v}_{2}\int
\dd\widehat{\boldsymbol {\sigma }}\,\Theta (\widehat{\boldsymbol
{\sigma }}\cdot
\mathbf{g})(\widehat{\boldsymbol {\sigma }}\cdot \mathbf{g})^{2}
\widehat{\boldsymbol {\sigma }}\widehat{\boldsymbol {\sigma }}  \int_{0}^{1} \dd x\; f^{(2)}\left[\mathbf{r}-x{\boldsymbol
{\sigma }},\mathbf{r}+(1-x) {\boldsymbol {\sigma }},\mathbf{v}_{1},\mathbf{v}_{2};t\right],
\label{2.11}
\end{equation}
\begin{equation}
{\bf q}^{c}({\bf r}, t)=\frac{1+\alpha}{4}m \sigma^{d}
\int \dd\mathbf{v}_{1}\int \dd\mathbf{v}_{2}\int
\dd\widehat{\boldsymbol {\sigma }}\,\Theta (\widehat{\boldsymbol
{\sigma }}\cdot
\mathbf{g})(\widehat{\boldsymbol {\sigma }}\cdot \mathbf{g})^{2}
({\bf G}\cdot\widehat{\boldsymbol {\sigma }})
\widehat{\boldsymbol {\sigma}}\int_{0}^{1}\dd x\; f^{(2)}\left[\mathbf{r}-
x{\boldsymbol{\sigma}},\mathbf{r}+(1-x)
{\boldsymbol {\sigma}},\mathbf{v}_{1},\mathbf{v}_{2};t\right],
\label{2.12}
\end{equation}
where $f^{(2)}$ is defined in Eq.\ \eqref{2.3} and  ${\bf G}=\frac{1}{2}({\bf V}_1+{\bf V}_2)$ is the velocity of the center of mass. Finally, the cooling rate is given by
\begin{equation}
\zeta({\bf r}, t) =\frac{\left(1-\alpha^{2}\right)}{4dnT} m \sigma^{d-1}\int \dd\mathbf{v}
_{1}\int \dd\mathbf{v}_{2}\int \dd\widehat{\boldsymbol {\sigma }}
\Theta (\widehat{\boldsymbol {\sigma }}\cdot
\mathbf{g})(\widehat{ \boldsymbol {\sigma }}\cdot
\mathbf{g})^{3}f^{(2)}(\mathbf{r}, \mathbf{r}+\boldsymbol {\sigma
},\mathbf{v}_{1},\mathbf{v}_{2};t). \label{2.13}
\end{equation}

Apart from the balance equations for the hydrodynamic fields $n$, $\mathbf{U}$, and $T$, to completely characterize the macroscopic state of the granular fluid one should also to derive the corresponding balance equations for the pressure tensor and the heat flux. These equations will be derived in the next section by considering an explicit form for the distribution function $f({\bf r},{\bf v},t)$.

\section{Grad's moment method}
\label{sec3}

Needless to say, to close the balance hydrodynamic equations \eqref{2.7}--\eqref{2.9} one needs to know the functional dependence of the momentum and heat fluxes and the cooling rate on the hydrodynamic fields $n$, $\mathbf{U}$ and $T$. A possible way of obtaining this dependence is solving the Enskog equation by means of the Chapman-Enskog method. \cite{CC70} This was the procedure followed in Refs.\ \onlinecite{GD99} and \onlinecite{L05} to determine $P_{ij}$, $\mathbf{q}$ and $\zeta$ to first order in the spatial gradients (NS hydrodynamic order). Here, a different procedure will be followed: Grad's method of moments. \cite{G49}
Although the method was originally devised to solve the Boltzmann equation for monatomic dilute gases, here it will be used to determine the NS transport coefficients of a granular dense fluid described by the inelastic Enskog equation \eqref{2.1}.

As mentioned in the Introduction, Grad's moment method is based on the expansion of the velocity distribution function in a complete set of orthogonal polynomials (generalized Hermite polynomials), the coefficients being the corresponding velocity moments. However, given that the (infinite) hierarchy of moment equations is not a closed set of equations, one has to truncate the above expansion after a certain order. After this truncation, the above hierarchy of moment equations becomes a closed set of coupled equations which can be solved. This allows one, for instance, to get the explicit forms of the NS transport coefficients. An interesting question is to assess the differences between the results derived from the Chapman-Enskog expansion and Grad's
moment method when only terms up to first order in the spatial gradients are retained in the constitutive equations for $P_{ij}$, $\mathbf{q}$, and $\zeta$.

In the application of the standard Grad moment method for a dense fluid, the retained moments are the hydrodynamic fields ($n$, $\mathbf{U}$, and $T$) plus the kinetic contributions to the irreversible momentum and heat fluxes ($P_{ij}^k-nT\delta_{ij}$ and $\mathbf{q}^k$). In the three-dimensional case ($d=3$), this implies that there are 13 moments involved in the form of the velocity distribution function $f$; hence this method is referred to as the 13-moment method. On the other hand, since we are interested in comparing the present results with those obtained for granular dense gases \cite{GD99,L05} from the Chapman-Enskog method, the full contracted moment of fourth order
\begin{equation}
\label{3.1}
c=\frac{8}{d(d+2)}\left[\frac{m^2}{4nT^2}\int\; \dd\mathbf{v}\; V^4  f(\mathbf{V})-\frac{d(d+2)}{4}\right]
\end{equation}
will be also included. The inclusion of the scalar field $c$ to the thirteen moments of mass
density, velocity, pressure tensor, and heat flux vector will allow us to make a close comparison with the previous Chapman-Enskog expressions derived for the cooling rate and the NS transport coefficients. \cite{GD99,L05}

Under the above conditions, the explicit form of the non-equilibrium distribution function $f({\bf r},{\bf v},t)$ can be written as
\begin{equation}
\label{3.2}
f(\mathbf{V})\to f_\text{M}(\mathbf{V}) \left[1 +\frac{m}{2nT^2}V_iV_j \Pi_{ij}+\frac{2}{d+2}\frac{m}{nT^2}\mathbf{S}(\mathbf{V})\cdot {\bf q}^k+\frac{c}{4}E(V)\right],
\end{equation}
where
\begin{equation}
\label{3.3}
f_\text{M}(\mathbf{V})=n\left(\frac{m}{2\pi T}\right)^{d/2}e^{-mV^2/2T}
\end{equation}
is the local equilibrium distribution function,
\begin{equation}
\label{3.3.1}
\mathbf{S}(\mathbf{V})=\left(
\frac{m V^2}{2T}-\frac{d+2}{2}\right){\bf V}, \quad E(V)=\left(\frac{mV^2}{2T}\right)^2-\frac{d+2}{2}
\frac{mV^2}{T}+\frac{d(d+2)}{4},
\end{equation}
and
\begin{equation}
\label{3.4}
\Pi_{ij}=P_{ij}^k-nT\delta_{ij}
\end{equation}
is the traceless part of the kinetic contribution to the pressure tensor. The coefficients appearing in each one of the velocity polynomials in Eq.\ \eqref{3.2} have been chosen by requiring that the basic hydrodynamics fields ($n$, $\mathbf{U}$, and $T$), the kinetic contributions to the pressure tensor and the heat flux vector, as well as the fourth contracted moment $c$ of the trial function \eqref{3.2} to be the same as those for the exact velocity distribution function $f$. As said in the Introduction, \emph{only} the velocity moments of the distribution function are present in Grad's solution. For this reason, the collisional contributions to the momentum and heat fluxes do not appear in the form \eqref{3.2} and they must be computed from their definitions \eqref{2.11} and \eqref{2.12} by replacing the (true) one-particle velocity distribution function $f$ by its Grad's approximation \eqref{3.2}. The collisional contributions to the fluxes have been determined in the Appendix \ref{appA}.

For elastic collisions ($\alpha=1$), the coefficient $c$ vanishes and so, one recovers the conventional form for the trial distribution $f$ in Grad's thirteen moment method. \cite{G49} Note that, for the sake of simplicity, in the fourteen-moment approximation \eqref{3.2}  some third-degree moments not included in the kinetic heat flux $\mathbf{q}^k$  have been left out. \cite{RRSC00,K11} The same can be said of the remaining polynomials of fourth order. The inclusion of the above moments would modify, for instance, the form of the cooling rate. However, as mentioned before, since we are interested in a theory with the same degree of accuracy as the one reported \cite{GD99,L05} by using the Chapman-Enskog method, only the pressure tensor and the heat flux, as well as the contracted fourth order moment \eqref{3.1}, will be included in the Grad's distribution function \eqref{3.2}.

\section{Navier-Stokes transport coefficients}
\label{sec4}

The form of the constitutive equations for the momentum and heat fluxes and the cooling rate in the NS order are given by Eqs.\ \eqref{1.2}--\eqref{1.4}, respectively. In this section, the NS transport coefficients and the cooling rate will be explicitly determined by using Grad's distribution \eqref{3.2}. For the sake of clarity, most of the technical details involved in these calculations are relegated to Appendices \ref{appA}, \ref{appB}, and \ref{appC} and only the final expressions for $\eta$, $\gamma$, $\kappa$, $\mu$, $\zeta_0$, and $\zeta_\text{U}$ will be displayed here. Let us consider first the momentum and heat fluxes.

\subsection{Momentum and heat fluxes}

The hydrostatic pressure $p=P_{ii}/d$ is given by
\begin{equation}
\label{4.1}
p=nT\left[1+2^{d-2}(1+\alpha)\phi \chi\right],
\end{equation}
where the solid volume fraction is defined by Eq.\ \eqref{1.7}.
To first order in the gradients, the pressure tensor $P_{ij}$ is given by Eq.\ \eqref{1.2}. As expected, while the shear viscosity $\eta$ has kinetic and collisional contributions, the bulk viscosity $\gamma$ has only a collisional contribution. The latter coefficient is
\begin{equation}
\label{4.5}
\gamma=\frac{2^{2d+1}}{\pi(d+2)}\phi^2 \chi (1+\alpha)\left(1-\frac{c_0}{32} \right)\eta_0,
\end{equation}
where
\begin{equation}
\label{4.6}
\eta_0=\frac{d+2}{8}\frac{\Gamma \left( \frac{d}{2}\right)}{\pi ^{\left( d-1\right) /2}}\sigma^{1-d}\sqrt{mT}
\end{equation}
is the low-density value of the NS shear viscosity in the elastic limit. The coefficient $c_0$ appearing in Eq.\ \eqref{4.5} characterizes the deviations of the distribution function $f$ from its Gaussian form in the homogeneous cooling state. It is given by
\begin{equation}
\label{4.20}
c_0=\frac{32(1-\alpha)(1-2\alpha^2)}{9+24d-\alpha(41-8d)+30(1-\alpha)\alpha^2}.
\end{equation}
This expression coincides with the one derived a few years ago by van Noije and Ernst. \cite{NE98}

The shear viscosity $\eta$ is given by
\begin{equation}
\label{4.4}
\eta=\eta_k\left[1+\frac{2^{d-1}}{d+2}\phi \chi (1+\alpha)\right]+\frac{d}{d+2}\gamma,
\end{equation}
where the subscript $k$ denotes the contributions to the transport coefficients coming from the kinetic parts of the fluxes. The kinetic part $\eta_k$ of the shear viscosity is
\begin{equation}
\label{4.7}
\eta_k=\frac{nT}{\nu_\eta-\frac{1}{2}\zeta_0}
\left[1-\frac{2^{d-2}}{d+2}(1+\alpha)
(1-3 \alpha)\phi \chi \right],
\end{equation}
where the collision frequency $\nu_\eta$ is
\begin{equation}
\label{4.8}
\nu_\eta=\frac{3\nu}{4d}\chi \left(1-\alpha+\frac{2}{3}d\right)(1+\alpha)
\left(1-\frac{c_0}{64}\right).
\end{equation}
Here, the nominal collision frequency $\nu$ is defined by
\begin{equation}
\label{4.8.1}
\nu=\frac{nT}{\eta_0}=\frac{8\pi ^{\left( d-1\right) /2}}{(d+2)\Gamma \left( \frac{d}{2}\right)}\sigma^{d-1}n\sqrt{\frac{T}{m}}.
\end{equation}
In Eq.\ \eqref{4.7}, the zeroth-order contribution $\zeta_0$ to the cooling rate is
\begin{equation}
\label{4.9} \zeta_0=\frac{d+2}{4d}(1-\alpha^2)\chi \left(1+\frac{3}{32}c_0\right)
\nu.
\end{equation}

To first order in the spatial gradients, the heat flux $\mathbf{q}$ is given by Eq.\ \eqref{1.3}. The thermal conductivity $\kappa$ and the coefficient $\mu$ can be written as
\begin{equation}
\label{4.10}
\kappa=\kappa_k\left[1+3\frac{2^{d-2}}{d+2}\phi \chi (1+\alpha)\right]+\frac{2^{2d+1}(d-1)}{(d+2)^2\pi}
\phi^2 \chi (1+\alpha)\left(1+\frac{7}{32} c_0 \right)\kappa_0,
\end{equation}
\begin{equation}
\label{4.11}
\mu=\mu_k\left[1+3\frac{2^{d-2}}{d+2}\phi \chi (1+\alpha)\right],
\end{equation}
where
\begin{equation}
\label{4.14}
\kappa_0=\frac{d(d+2)}{2(d-1)}\frac{\eta_0}{m}
\end{equation}
is the low-density value of the thermal conductivity of an elastic gas. In addition, the kinetic parts $\kappa_k$ and $\mu_k$ are
\begin{equation}
\label{4.12}
\kappa_k=\frac{d-1}{d}\kappa_0\nu
\left(\nu_\kappa-2\zeta_0\right)^{-1}\left\{1+c_0+3\frac{2^{d-3}}{d+2}\phi \chi(1+\alpha)^2\left[2\alpha-1+\frac{c_0}{2}(1+\alpha)\right]\right\},
\end{equation}
\begin{eqnarray}
\label{4.13} \mu_k&=&\frac{\kappa_0 T\nu}{n}\left(\nu_\kappa-\frac{3}{2}\zeta_0\right)^{-1}\left\{\zeta_0^*\kappa_k^*\left(1+\phi\partial_\phi\ln
\chi\right)
+\frac{d-1}{2d}c_0+3\frac{2^{d-2}(d-1)}{d(d+2)}\phi \chi
(1+\alpha)\right. \nonumber\\
& &\left. \times \left(1+\frac{1}{2}\phi\partial_\phi\ln
\chi\right)\left[\alpha(\alpha-1)+\frac{c_0}{12}(10+2d-3\alpha+3\alpha^2)\right]\right\}.
\end{eqnarray}
In Eqs.\ \eqref{4.12} and \eqref{4.13}, $\zeta_0^*\equiv \zeta_0/\nu$, $\kappa_k^*\equiv \kappa_k/\kappa_0$, and the collision frequency $\nu_\kappa$ is
\begin{equation}
\label{4.15}
\nu_\kappa=
\frac{1+\alpha}{d}\nu\chi \left[\frac{d-1}{2}+\frac{3}{16}(d+8)
(1-\alpha)+\frac{4+5d-3(4-d)\alpha}{1024}c_0\right].
\end{equation}

\subsection{Fourth moment and cooling rate}

In the first order of the spatial gradients, the fourth moment $c$ is given by
\begin{equation}
\label{4.19.1}
c=c_0+c_1 \nabla \cdot \mathbf{U},
\end{equation}
where the coefficient $c_0$ is given by Eq.\ \eqref{4.20}. The first-order contribution $c_1$ is
\begin{equation}
\label{4.17}
c_1=-\frac{4\lambda-2^{d+1}(d+2)\chi \phi (1+\alpha)\left(1-3\alpha\right)\left(
1+\frac{c_0}{2}\right)}{\nu_\zeta+\frac{19}{16}d(d+2)^2\chi (1-\alpha^2)},
\end{equation}
where
\begin{equation}
\label{4.18}
\lambda=2^{d-3}\phi \chi (1+\alpha)\left\{5+4d(1-3\alpha)-9\alpha+3\alpha^2-15\alpha^3-
\frac{c_0}{4}\left[15\alpha^3-3\alpha^2+3(4d+15)\alpha-(20d+1)\right]\right\},
\end{equation}
\begin{equation}
\label{4.19}
\nu_\zeta=-\frac{d+2}{32}\chi (1+\alpha)\left[32(d-1)+3(1-\alpha)(10\alpha^2+10d+39)\right].
\end{equation}

Finally, to first order in the gradients, the cooling rate $\zeta$ is expressed by Eq.\ \eqref{1.4}, where $\zeta_0$ is given by Eq.\ \eqref{4.9} and
\begin{equation}
\label{4.16bis}
\zeta_\text{U}=-\frac{3}{d}\chi (1-\alpha^2)\left(2^{d-2}\phi-\frac{d+2}{128}c_1 \right).
\end{equation}

It must be noted that Navierû-Stokes hydrodynamics retains terms up through second order in the spatial gradients. Since the cooling rate $\zeta$ is a scalar, its most general form at this order is
\begin{eqnarray}
\label{4.20.1}
\zeta&=&\zeta_0+\zeta_\text{U} \nabla \cdot \mathbf{U}+\zeta_{n}\nabla^2 n+\zeta_{T}\nabla^2 T
+\zeta_{nn}(\nabla n)^2+\zeta_{TT}(\nabla T)^2\nonumber\\
& & +\zeta_{n T}\nabla n \cdot \nabla T+
\zeta_{1,uu}\partial_i U_j \partial_i U_j+\zeta_{2,uu}\partial_j U_i \partial_i U_j.
\end{eqnarray}
The first two second-order terms $\zeta_n$ and $\zeta_T$ have been determined for dilute granular gases by Brey \emph{et al.} \cite{BDKS98} while \emph{all} the set of coefficients $\left\{\zeta_n, \zeta_T, \zeta_{nn}, \zeta_{nT}, \zeta_{1,uu}, \zeta_{2,uu} \right\}$ have been computed for granular gases of viscoelastic particles by Brilliantov and P\"oschel. \cite{BP03} The evaluation of the above set of coefficients for dense gases is a quite intricate problem. In fact, to the best of my knowledge, no explicit results for these coefficients have been reported for granular dense gases. However, it has been shown for dilute gases that the contributions of the second-order terms to the cooling rate $\zeta$ are negligible \cite{BDKS98,BP03} as compared with the corresponding zeroth-order contribution $\zeta_0$ (the first-order contribution $\zeta_\text{U}$ vanishes for dilute gases). It is assumed here that the same holds in the dense case and so, for practical applications these second-order contributions can be in principle neglected in the Navier-Stokes hydrodynamic equations.

\subsection{Dilute granular gas}

An interesting particular case corresponds to the low-density limit ($\phi=0$). In this case, $\gamma=\zeta_\text{U}=0$ and the NS transport coefficients become
\begin{equation}
\label{4.21}
\eta=\frac{nT}{\nu_\eta-\frac{1}{2}\zeta_0},
\end{equation}
\begin{equation}
\label{4.22}
\kappa=\frac{d-1}{d}\kappa_0\nu \frac{1+c_0}{\nu_\kappa-2\zeta_0},
\end{equation}
\begin{equation}
\label{4.23}
\mu=\frac{T}{n}\frac{\zeta_0\kappa
+\frac{d-1}{2d}\kappa_0\nu c_0}{\nu_\kappa-\frac{3}{2}\zeta_0}.
\end{equation}
Here, $\nu_\eta$, $\zeta_0$, and $\nu_\kappa$ are given by Eqs.\ \eqref{4.8}, \eqref{4.9}, and \eqref{4.15}, respectively, with $\chi=1$. Equations \eqref{4.21}--\eqref{4.23} agree with those recently obtained \cite{RGD12} from the Boltzmann equation. In addition, the results reported here coincide with those derived by Kremer and Marques \cite{K11} for a dilute granular gas when one neglects the cooling effects on the granular temperature (i.e., when one formally takes $\zeta_0=0$ in Eqs.\ \eqref{4.21}--\eqref{4.23}).

\section{Comparison with other results}
\label{sec5}

\begin{figure}
\includegraphics[width=0.45 \columnwidth,angle=0]{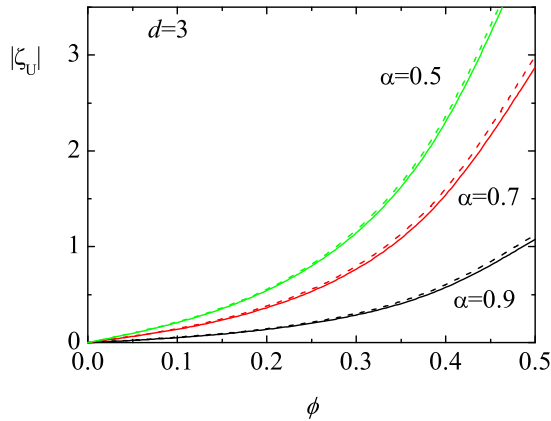}
\caption{(color online) Plot of the magnitude of the first-order contribution $\zeta_\text{U}$ to the cooling rate versus
the solid volume fraction $\phi$ for hard spheres ($d=3$) and three different values of the coefficient of restitution $\alpha$: $\alpha=0.9$, $\alpha=0.7$, and $\alpha=0.5$. The solid lines are the results obtained from Grad's method, while the dashed lines correspond to those obtained from the Chapman-Enskog method.
\label{fig1}}
\end{figure}

The results derived in the preceding section provide the expressions of the NS transport coefficients and the cooling rate obtained by solving the \emph{inelastic} Enskog equation by means of a fourteen moment method. More specifically, the bulk ($\gamma$) and shear ($\eta$) viscosities are given by Eqs.\ \eqref{4.5} and \eqref{4.4}, respectively, the thermal conductivity $\kappa$ is given by Eqs.\ \eqref{4.10} and \eqref{4.12}, the coefficient $\mu$ is given by Eqs.\ \eqref{4.11} and \eqref{4.13}, and the cooling rate coefficient $\zeta_\text{U}$ is given by Eq.\ \eqref{4.16bis}. It is quite apparent that the reduced forms $\eta^*$, $\gamma^*$, $\kappa^*$, and $\mu^*$ of the above transport coefficients [see Eqs.\ \eqref{1.5} and \eqref{1.6}] present a complex dependence on both the coefficient of restitution $\alpha$ and the solid volume fraction $\phi$. In order to get their explicit forms, the dependence of the pair correlation function $\chi$ on $\phi$ must be chosen. In the three-dimensional case ($d=3$), a good approximation for $\chi$ is given by the Carnahan-Starling expression \cite{CS69}
\begin{equation}
\label{5.1}
\chi(\phi)=\frac{1-\frac{1}{2}\phi}{(1-\phi)^3},
\end{equation}
while for hard disks ($d=2$), $\chi$ can be approximated by \cite{H75}
\begin{equation}
\label{5.2}
\chi(\phi)=\frac{1-\frac{7}{16}\phi}{(1-\phi)^2}.
\end{equation}

A comparison with the results \cite{GD99} obtained from the Chapman-Enskog method for inelastic hard spheres ($d=3$) shows that the expressions of the NS transport coefficients $\eta$, $\gamma$, $\kappa$, and $\mu$ are the same as those obtained here from Grad's moment method. In the case of arbitrary number of dimensions $d$, the above expressions also agree with those obtained first by Lutsko \cite{L05} and more recently by the author of the present paper. \cite{G12,note} On the other hand, the first order contribution $\zeta_\text{U}$ to the cooling rate [see Eq.\ \eqref{4.16bis}] is different from the one derived in the Chapman-Enskog theory. However, a study of the dependence of $\zeta_\text{U}$ on $\phi$ and $\alpha$ shows that the results obtained from both methods are very similar. To illustrate these differences, Fig.\ \ref{fig1} shows the magnitude $|\zeta_\text{U}|$ versus the solid volume fraction $\phi$ for three different values of the coefficient of restitution ($\alpha=0.9$, 0.7 and 0.5) in the case of hard spheres ($d=3$). We observe that the Chapman-Enskog and Grad predictions for $|\zeta_\text{U}|$ are indistinguishable in all the range of values of $\phi$ analyzed. A similar behavior is also found for a two-dimensional system. Consequently, we can conclude that both methods essentially yield the same results in the NS regime.

\begin{figure}
\begin{tabular}{lr}
\resizebox{7.2cm}{!}{\includegraphics{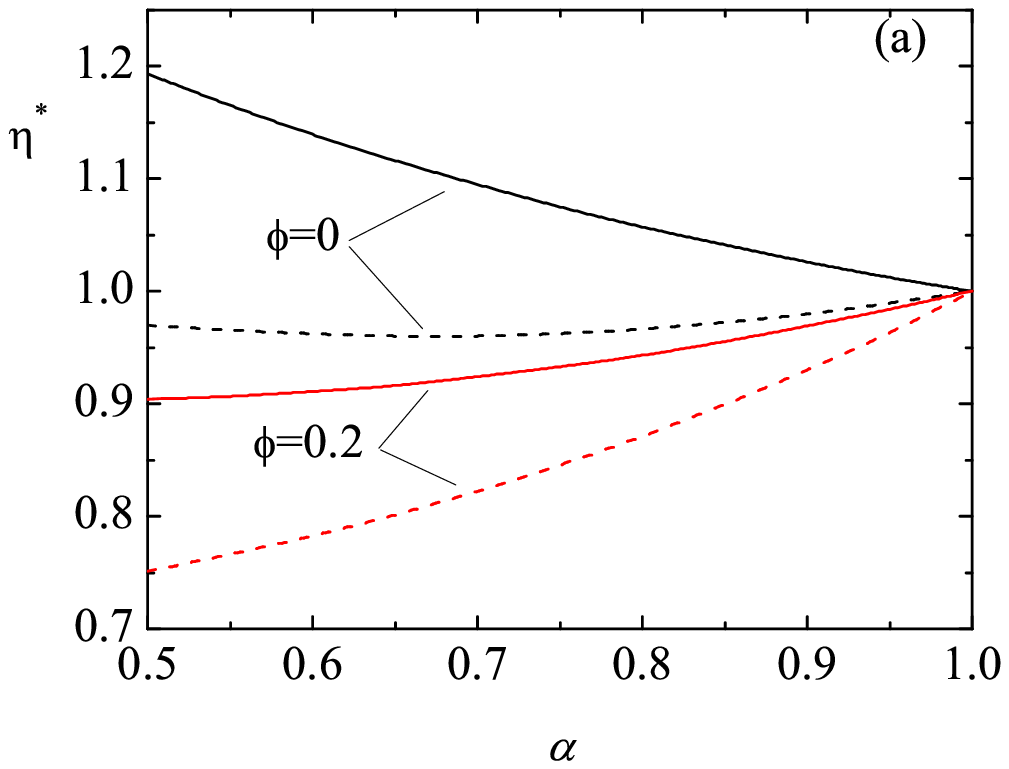}}
&\resizebox{7cm}{!}{\includegraphics{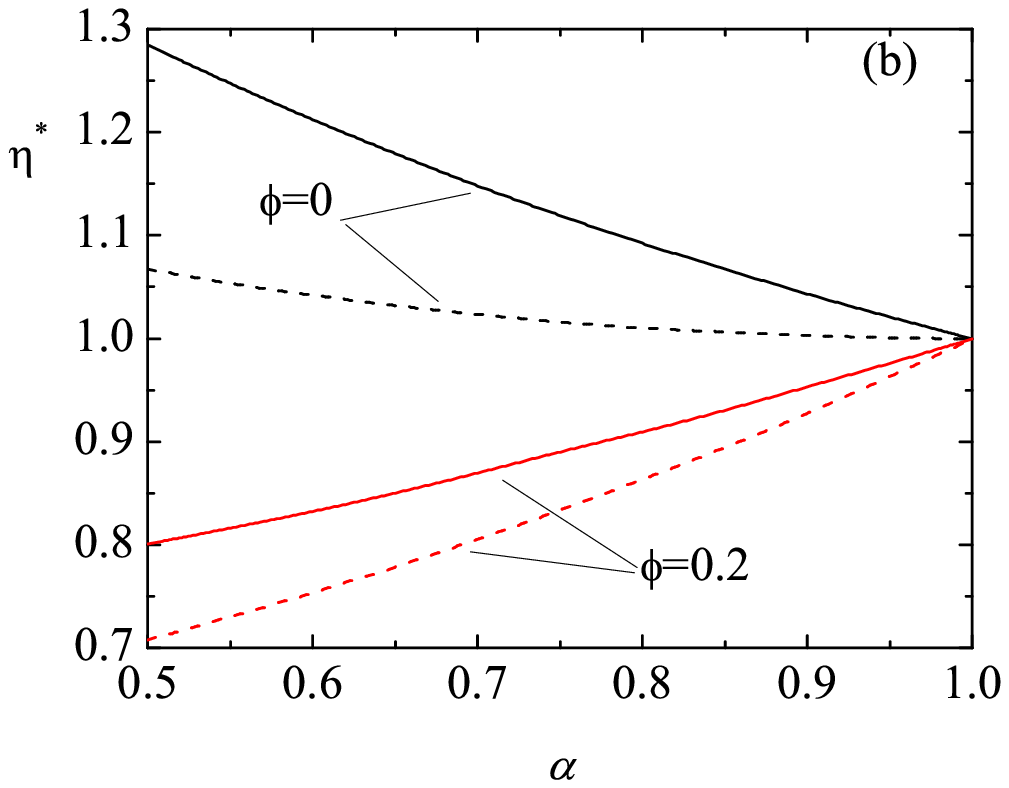}}
\end{tabular}
\caption{(color online) Plot of the reduced shear viscosity $\eta^*(\alpha,\phi)=\eta(\alpha,\phi)/\eta(1,\phi)$ versus the coefficient of restitution $\alpha$ for two different values of the solid volume fraction ($\phi=0$ and $\phi=0.2$) in the cases of hard disks (a) and hard spheres (b). The solid lines are the results derived here, while the dashed lines are the results obtained by Jenkins and Richman. \cite{JR85a,JR85b} \label{fig2}}
\end{figure}
\begin{figure}
\begin{tabular}{lr}
\resizebox{7.2cm}{!}{\includegraphics{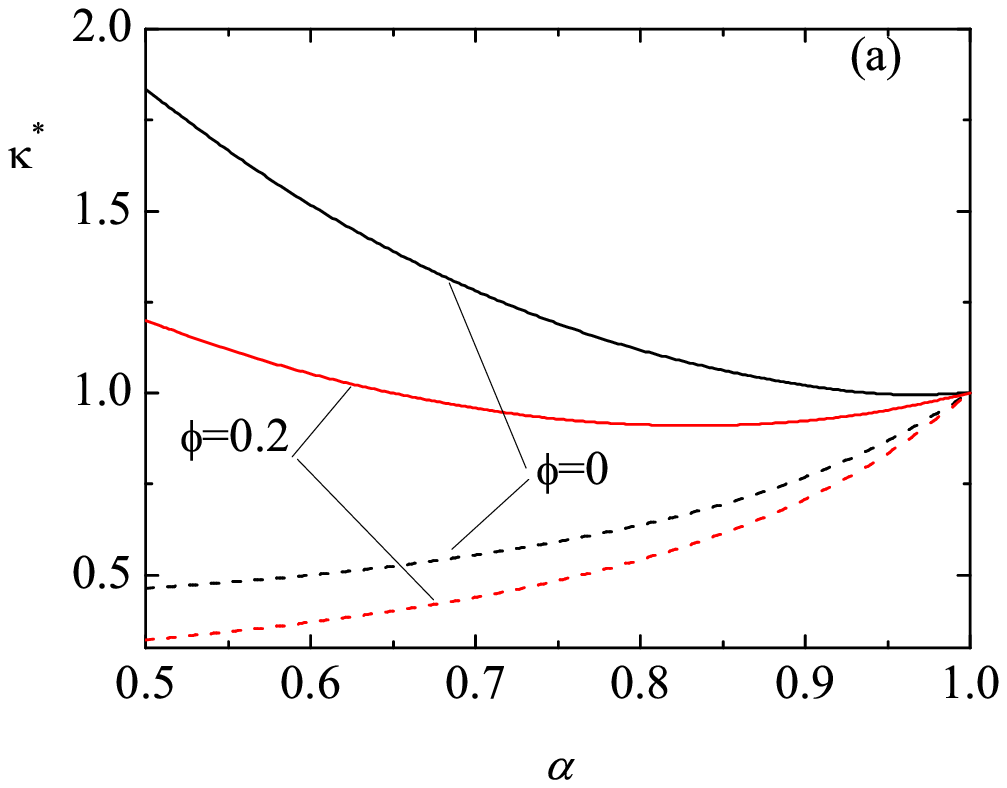}}
&\resizebox{7cm}{!}{\includegraphics{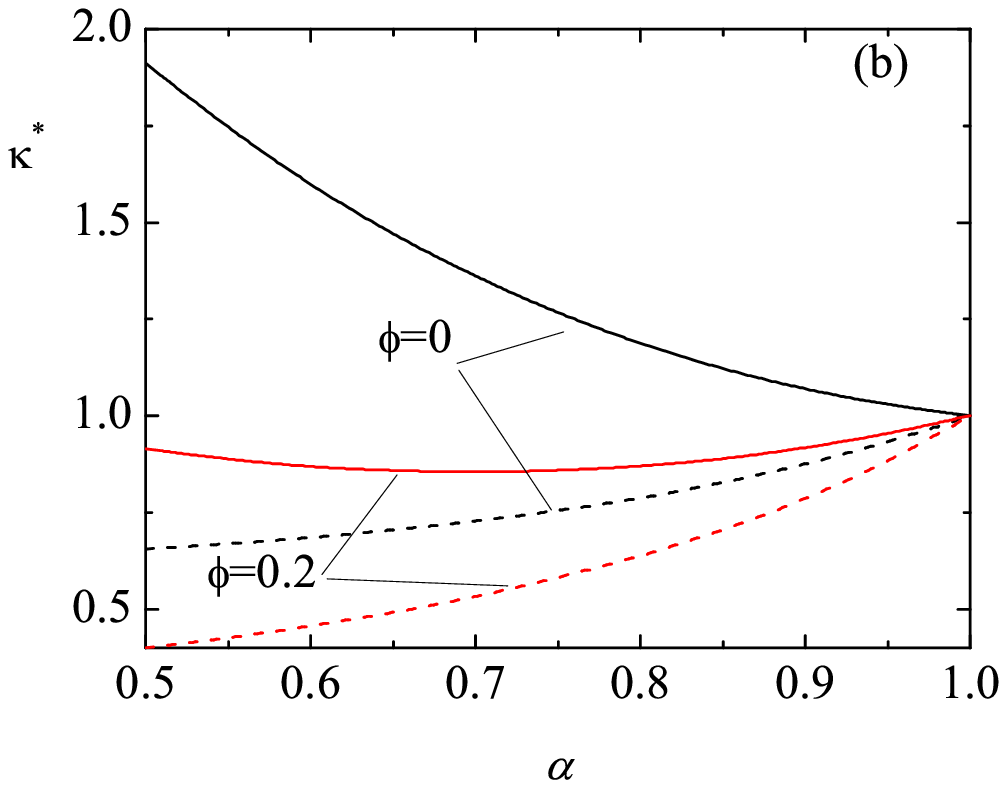}}
\end{tabular}
\caption{(color online) Plot of the reduced thermal conductivity $\kappa^*(\alpha,\phi)=\kappa(\alpha,\phi)/\kappa(1,\phi)$ versus the coefficient of restitution $\alpha$ for two different values of the solid volume fraction ($\phi=0$ and $\phi=0.2$) in the cases of hard disks (a) and hard spheres (b). The solid lines are the results derived here, while the dashed lines are the results obtained by Jenkins and Richman. \cite{JR85a,JR85b} \label{fig3}}
\end{figure}
\begin{figure}
\begin{tabular}{lr}
\resizebox{7.1cm}{!}{\includegraphics{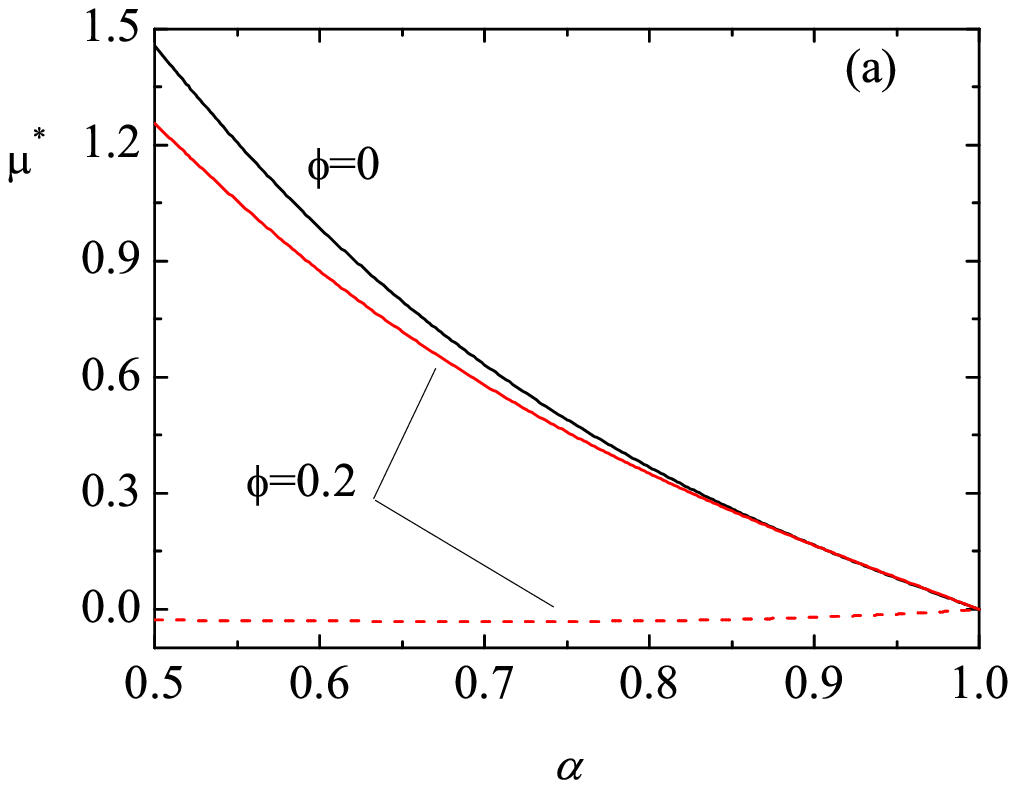}}
&\resizebox{7cm}{!}{\includegraphics{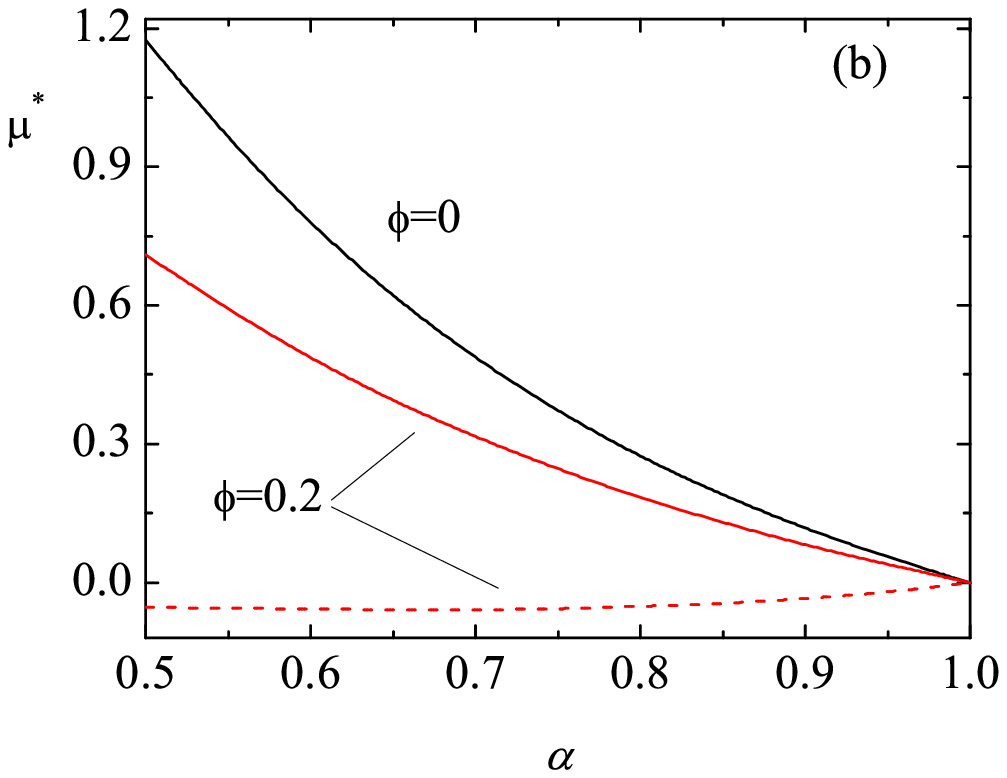}}
\end{tabular}
\caption{(color online) Plot of the reduced coefficient $\mu^*(\alpha,\phi)=n\mu(\alpha,\phi)/T\kappa(1,\phi)$ versus the coefficient of restitution $\alpha$ for two different values of the solid volume fraction ($\phi=0$ and $\phi=0.2$) in the cases of hard disks (a) and hard spheres (b). The solid lines are the results derived here, while the dashed lines are the results obtained by Jenkins and Richman. \cite{JR85a,JR85b} Note that the coefficient $\mu$ vanishes in the Jenkins--Richman theory for a dilute gas ($\phi=0$). \label{fig4}}
\end{figure}
\begin{figure}
\includegraphics[width=0.45 \columnwidth,angle=0]{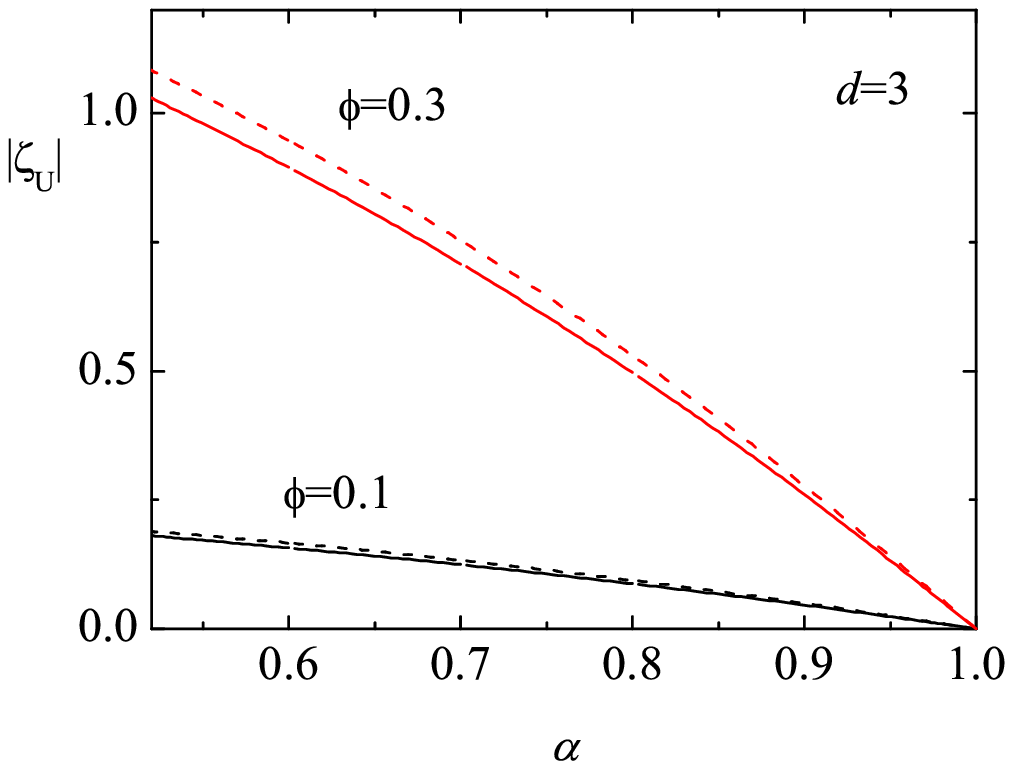}
\caption{(color online) Plot of the magnitude of the first-order contribution $\zeta_\text{U}$ to the cooling rate versus
the coefficient of restitution $\alpha$ for hard spheres ($d=3$) and two different values of the solid volume fraction $\phi$: $\phi=0.1$ and $\phi=0.3$. The solid lines are the results obtained here, while the dashed lines correspond to those obtained by Jenkins and Richman. \cite{JR85b}
\label{fig5}}
\end{figure}

Apart from comparing the present results with those derived from another different method (Chapman-Enskog expansion), it is quite instructive to make a comparison between the results of this paper with those previously reported in the literature by employing a different version of Grad's method. As mentioned in the Introduction, Grad's 13-moment method was already used many years ago by Jenkins and Richman to determine the NS transport coefficients of a dense gas of inelastic hard disks \cite{JR85a} and spheres. \cite{JR85b} Their explicit results are displayed in Appendix \ref{appD} for the sake of completeness. As usual in the conventional Grad's method, these authors did not include the fourth order polynomial $E(V)$ in the trial solution \eqref{3.2} and so the coefficients $c_0$ and $c_1$ vanish in their approximation.

A careful comparison between the results displayed in Sec.\ \ref{sec4} with those provided in Appendix \ref{appD} shows that the expressions for $\eta$, $\gamma$, $\kappa$, and $\mu$ are different in both approaches. The differences are essentially due to the assumptions made in Refs.\ \onlinecite{JR85a} and \onlinecite{JR85b} since the authors neglect the time dependence of temperature due to collisional cooling (which is formally equivalent to take $\zeta_0=0$) and, as said before, they do not include the fourth degree polynomial $E(V)$ in Grad's solution \eqref{3.2} (which is equivalent to take $c_0=c_1=0$). It is important to remark that, while the latter simplification is in general not relevant (except for very small values of the coefficient of restitution), the former assumption turns out to be quite significant beyond the quasielastic limit (i.e., for finite values of the coefficient of restitution).

In order to illustrate the quantitative differences between the Jenkins-Richman theory \cite{JR85a,JR85b} and the results displayed in Sec.\ \ref{sec4}, Figs.\ \ref{fig2}, \ref{fig3} and \ref{fig4} show the $\alpha$-dependence of the reduced transport coefficients $\eta^*(\alpha,\phi)\equiv \eta(\alpha,\phi)/\eta(1,\phi)$, $ \kappa^*(\alpha,\phi)\equiv \kappa(\alpha,\phi)/\kappa(1,\phi)$, and $
\mu^*(\alpha,\phi)\equiv n\mu(\alpha,\phi)/T\kappa(1,\phi)$, respectively. Two different values of the solid volume fraction $\phi$ have been considered: a dilute gas ($\phi=0$) and a moderately dense gas ($\phi=0.2$). Moreover, for the sake of completeness, the above reduced transport coefficients have been plotted for disks ($d=2$) and spheres ($d=3$).

It is apparent from Fig.\ \ref{fig2} that the qualitative dependence of the shear viscosity on dissipation is relatively well captured by the Jenkins-Richman theory, especially in the three-dimensional case. In this case ($d=3$), both Grad's solutions predict that while $\eta^*$ is an increasing function of dissipation for dilute gases, the opposite happens at moderate densities. However, at a more quantitative level, there are important discrepancies between both theories, especially as the coefficient of restitution $\alpha$ decreases. The differences for the heat flux transport coefficients are much more significant than those observed for the shear viscosity. In the case of the thermal conductivity, Fig.\ \ref{fig3} shows that both theories predict different qualitative behavior for a dilute gas ($\phi=0$): while the present theory shows that the latter coefficient increases with decreasing $\alpha$, the opposite happens in the Jenkins-Richman theory. As the density increases, although the differences between both theories are smaller than that of a low-density gas, they are still large for finite dissipation, especially in the case of hard disks. As expected, the discrepancies are much more important for the coefficient $\mu$ (which vanishes in the elastic limit). In particular, the coefficient $\mu$ becomes negative in the Jenkins-Richman theory [see Eqs.\ \eqref{d3} and \eqref{d7}] for inelastic dense gases ($\phi \neq 0$), but its magnitude is practically zero. This drawback is not present in our results since $\mu$ is always positive for any value of $\alpha$ and $\phi$. In addition, according to the results presented here, although the magnitude of $\mu$ is in general smaller than that of the thermal conductivity $\kappa$, we observe that the influence of $\mu$ on the heat transport could not be considered as negligible as the degree of dissipation increases.

Finally, the dependence of the magnitude of the first order contribution to the cooling rate $|\zeta_\text{U}|$ on the coefficient of restitution is plotted in Fig.\ \ref{fig5} for hard spheres and two values of the solid volume fraction. Note that the results derived by Jenkins and Richman \cite{JR85b} for $\zeta_\text{U}$ coincide with those obtained here when one neglects the coefficient $c_1$ defined by Eq.\ \eqref{4.17}. Figure \ref{fig5} shows that the agreement between both theories is quite good, indicating that the influence of the coefficient $c_1$ on $|\zeta_\text{U}|$ is relatively small. Thus, in contrast to the behavior observed in some of the previous transport coefficients, the expression for $\zeta_\text{U}$ obtained by Jenkins and Richman \cite{JR85b} can be considered as reliable, even for finite values of dissipation.

\section{Conclusions}
\label{sec6}

In this paper, the NS transport coefficients of a granular fluid have been determined by solving the \emph{inelastic} Enskog equation by means of Grad's moment method. As in previous works on dilute granular gases, \cite{RGD12,RRSC00,K11} the solution proposed here differs from the conventional thirteen moment method by the inclusion of the full contracted fourth moment $c$ defined by Eq.\ \eqref{3.1}. To first order in the spatial gradients, this moment can be written as $c=c_0+c_1\nu^{-1}\nabla \cdot \mathbf{U}$, where $\nu$ is an effective collision frequency defined by Eq.\ \eqref{4.8.1}, $c_0$ is given by Eq.\ \eqref{4.20} and the expression of $c_1$ is given by Eq.\ \eqref{4.17}. While the fourth cumulant $c_0$ takes into account the contributions to the transport coefficients coming from non-Gaussian corrections  to the homogenous cooling state, the presence of $c_1$ gives rise to a new first-order contribution to the cooling rate. The coefficient $c_0$ vanishes for elastic collisions while $c_1=0$ in the limits of elastic dense spheres ($\alpha=1$ but $\phi\neq 0$) and of dilute inelastic spheres ($\alpha \neq 1$ but $\phi=0$). Thus, although both coefficients are different from zero for inelastic dense fluids, their impact on the NS transport coefficients is in general quite small and, consequently, one can neglect their contributions to the NS transport coefficients even for finite degree of dissipation.

On the other hand, in contrast to the previous works \cite{JR85a,JR85b,RRSC00,K11}  for granular gases from Grad's moment method,  the present results take also into account the time dependence of the granular temperature due to cooling effects. This dependence is accounted for by the zeroth-order cooling rate $\zeta_0$ given by Eq.\ \eqref{4.9}. In fact, if one neglects the small corrections due to $c_0$ and $c_1$, the present results agree with the previous ones for dense gases \cite{JR85a,JR85b} when one takes $\zeta_0=0$ in the expressions displayed in Sec.\ \ref{sec4}. The quantitative variation of the (reduced) transport coefficients on both the coefficient of restitution and density has been widely illustrated in Figs.\ \ref{fig2}--\ref{fig4}. The comparison with the Jenkins-Richman theory \cite{JR85a,JR85b}  clearly shows that in general the influence of $\zeta_0$ on $\eta^*$ (see Fig.\ \ref{fig2}), $\kappa^*$ (see Fig.\ \ref{fig3}), and $\mu^*$ (see Fig.\ \ref{fig4}) is significant and so the cooling effects on granular temperature  cannot be neglected beyond the quasielastic limit ($\alpha \simeq 1$).

Furthermore, the expressions obtained here for the NS transport coefficients and the cooling rate agree completely with those obtained several years ago from the Chapman-Enskog expansion by considering the first Sonine approximation. \cite{GD99,L05} As for ordinary gases, \cite{C90,CC70} this agreement shows the equivalence between both approximate methods to solve the Enskog equation for granular dense gases in the NS regime. It must be remarked that the inclusion of more velocity moments (for instance, all the third and fourth degree velocity moments) would change the final results since for instance there would be likely additional contributions to the cooling rate. Nevertheless, for practical purposes, the inclusion of those new terms in Grad's solution \eqref{3.2} makes analytic calculations much more difficult since higher order collision integrals should be evaluated to compute the new contributions to the momentum and heat fluxes.

The derivation of explicit expressions for the transport coefficients is perhaps one of the most important challenges of granular gas research. The theoretical results reported in this paper cover part of this challenge, at least in the NS domain. Nevertheless, the present results have some restrictions. First, although the Enskog equation retains spatial correlations arising from volume exclusion effects, it still assumes uncorrelated particle velocities (molecular chaos hypothesis). Therefore, it is expected that the results reported here only apply to \emph{moderate} densities (solid volume fraction typically smaller than or equal to 0.25). However, in spite of this limitation, there is substantial evidence in the literature \cite{RET} on the reliability of the Enskog kinetic theory to accurately describe macroscopic properties (such as transport coefficients) for a wide range of densities and/or collisional dissipation. Another important limitation is the accuracy of the results obtained here for quite extreme values of dissipation. As mentioned in the Introduction, although the approximate expressions for the NS transport coefficients displayed in Sec.\ \ref{sec4} compare in general quite well with computer simulations, \cite{MonteCarlo} there are significant discrepancies for the heat flux transport coefficients for small values of $\alpha$ (say, for instance, $\alpha \lesssim 0.7$). Therefore, the reliability of the expressions \eqref{4.10} and \eqref{4.11} for the thermal conductivity $\kappa$ and the coefficient $\mu$, respectively, can be questionable for this range of small values of the coefficient of restitution. In the context of the Chapman-Enskog solution, the above discrepancies between theory and simulation can be in part mitigated \cite{GSM07,Peter} when one uses the homogeneous cooling state distribution instead of the Maxwellian distribution as the weight function in the corresponding first-Sonine approximation. Finally, the evaluation of the second-order contributions (defined in Eq.\ \eqref{4.20.1}) to the cooling rate  from Grad's moment method could be a possible future work. This would allow us to assess their impact on the cooling rate for finite densities. In addition, the knowledge of these second-order terms along with the transport coefficients derived in this paper would provide us the complete set of transport coefficients needed to solve the nonlinear Navier-Stokes hydrodynamic equations.

\acknowledgements

I am grateful to Dr. Andr\'es Santos for a critical reading of the manuscript. The present work has been supported by the Ministerio de Educaci\'on y Ciencia (Spain) through grant No. FIS2010-16587, partially financed by
FEDER funds and by the Junta de Extremadura (Spain) through Grant No. GR10158.

\appendix
\section{Collisional contributions to the fluxes}
\label{appA}

The collisional transfer contributions to the pressure tensor and heat flux are determined from Eqs.\ \eqref{2.11} and \eqref{2.12}, respectively. In order to get these collisional contributions to first order in gradients, it is convenient to express first the contracted fourth moment $c$ to linear order. Thus, based on previous results derived in the NS order for \emph{dense} gases, \cite{GD99} it is expected that in the linear theory the trial distribution function \eqref{3.2} must depend on the divergence of flow velocity through the moments acting as coefficients. Given that the tensor $\Pi_{ij}$ is traceless, the only contribution proportional to $\nabla \cdot \mathbf{U}$ in Grad's solution \eqref{3.2} comes from the fourth moment $c$. This necessarily implies (as we will show later in Appendix \ref{appC}) that the coefficient $c$ can be decomposed as
\begin{equation}
\label{3.5}
c=c_0+c_1 \nu^{-1} \nabla \cdot \mathbf{U},
\end{equation}
where $\nu$ is defined by Eq.\ \eqref{4.8.1}. According to the decomposition \eqref{3.5}, while the coefficient $c_0$  characterizes the deviations of $f$ from its Gaussian form in the homogenous cooling state, the coefficient $c_1$ gives rise to a new first-order contribution to the cooling rate $\zeta$. This new contribution was not accounted for in previous results \cite{JR85a,JR85b} based on Grad's method.

Let us evaluate now the collisional contributions. Consider first the pressure tensor which becomes up to first order in the gradients
\begin{eqnarray}
\label{a1}
P_{ij}^c&=&\frac{1+\alpha}{4}m\sigma^{d}\chi \int\; \dd\mathbf{V}_1 \int\; \dd\mathbf{V}_2
\int \dd\widehat{\boldsymbol{\sigma}}\,\Theta (\widehat{{\boldsymbol {\sigma}}}
\cdot {\bf g})(\widehat{\boldsymbol {\sigma }}\cdot {\bf g})^2 \widehat{\sigma}_i\widehat{\sigma}_j\left[f_0(\mathbf{V}_1)f_0(\mathbf{V}_2)+
f_1(\mathbf{V}_1)f_0(\mathbf{V}_2)\right.
\nonumber\\
& &\left.+f_0(\mathbf{V}_1)f_1(\mathbf{V}_2)-\frac{1}{2}f_0(\mathbf{V}_2)
\widehat{{\boldsymbol {\sigma}}}\cdot \nabla f_0(\mathbf{V}_1)+
\frac{1}{2}f_0(\mathbf{V}_1)
\widehat{{\boldsymbol {\sigma}}}\cdot \nabla f_0(\mathbf{V}_2)\right],
\end{eqnarray}
where
\begin{equation}
\label{a2}
f_0(\mathbf{V})=f_\text{M}(\mathbf{V}) \left(1 +\frac{c_0}{4}E(V)\right),
\end{equation}
is the part of the trial distribution \eqref{3.2} of zeroth-order in spatial gradients and
\begin{equation}
\label{a3}
f_1(\mathbf{V})=f_\text{M}(\mathbf{V})\left[
\frac{m}{2nT^2}V_iV_j \Pi_{ij}+\frac{2}{d+2}\frac{m}{nT^2}\mathbf{S}(\mathbf{V})\cdot {\bf q}^k
+\frac{c_1}{4}E(V)\nu^{-1}\nabla \cdot \mathbf{U}\right]
\end{equation}
is the part of the trial distribution \eqref{3.2} of first-order in spatial gradients. Upon writing Eqs.\ \eqref{a1}--\eqref{a3} use has been made of Eq.\ \eqref{3.5}. The integration over solid angle in Eq.\ \eqref{a1} yields
\begin{eqnarray}
\label{a4}
P_{ij}^c&=&\frac{1+\alpha}{4}\frac{B_2}{d+2}m\sigma^{d}\chi \int\; \dd\mathbf{V}_1 \int\; \dd\mathbf{V}_2 \left[f_0(\mathbf{V}_1)f_0(\mathbf{V}_2)+2f_1(\mathbf{V}_1)f_0(\mathbf{V}_2)\right]
\left(2g_ig_j+g^2\delta_{ij}\right)
\nonumber\\
& &-\partial_k U_\ell \frac{1+\alpha}{4}\frac{B_3}{d+3}m\sigma^{d+1}\chi \int\; \dd\mathbf{V}_1 \int\; \dd\mathbf{V}_2 \;f_0(\mathbf{V}_1)\frac{\partial f_0(\mathbf{V}_2)}{\partial V_{2\ell}}g^{-1}\left[g_ig_jg_k+g^2\left(g_k \delta_{ij}+g_i \delta_{jk}+g_j \delta_{ik}\right)\right],
\end{eqnarray}
where the coefficients $B_n$ are defined by\cite{NE98}
\begin{equation}
B_{n}\equiv \int \dd\widehat{\boldsymbol{\sigma}}\,\Theta (\widehat{{\boldsymbol {\sigma}}}
\cdot {\bf g})(\widehat{\boldsymbol {\sigma }}\cdot {\bf g})^n=\pi ^{\left( d-1\right) /2}\frac{\Gamma \left(
\frac{n+1}{2} \right) }{\Gamma \left( \frac{n+d}{2}\right) }.
\label{a5}
\end{equation}
The expression \eqref{a4} can be more explicitly written when one takes into account the forms of $f_0$ and $f_1$. After some algebra, the result is
\begin{eqnarray}
\label{a6}
P_{ij}^c&=&B_2\frac{1+\alpha}{2} n \sigma^d \chi nT+\frac{B_2}{d+2}(1+\alpha) n \sigma^d \chi \Pi_{ij}
\nonumber\\
& &+\partial_k U_\ell \frac{1+\alpha}{4}\frac{B_3}{d+3}m\sigma^{d+1}\chi \int\; \dd\mathbf{V}_1 \int\; \dd\mathbf{V}_2 \;f_0(\mathbf{V}_1)
f_0(\mathbf{V}_2)\frac{\partial}{\partial V_{2\ell}}\left\{g^{-1}\left[g_ig_jg_k+g^2\left(g_k \delta_{ij}+g_i \delta_{jk}+g_j \delta_{ik}\right)\right]\right\}\nonumber\\
&=&B_2\frac{1+\alpha}{2} n \sigma^d \chi nT+\frac{B_2}{d+2}(1+\alpha) n \sigma^d \chi \Pi_{ij}
\nonumber\\
& &-B_3 \frac{d+1}{4d^2}m\sigma^{d+1}\int\; \dd\mathbf{V}_1 \int\; \dd\mathbf{V}_2\; f_0(\mathbf{V}_1)
f_0(\mathbf{V}_2)g\left[\frac{d}{d+2}\left(\partial_j U_i+\partial_i U_j-\frac{2}{d}\delta_{ij}
\nabla \cdot \mathbf{U}\right)+\delta_{ij}\nabla \cdot \mathbf{U}\right]\nonumber\\
&=&B_2\frac{1+\alpha}{2} n \sigma^d \chi n T+\frac{B_2}{d+2}(1+\alpha) n \sigma^d \chi \Pi_{ij}
\nonumber\\
& &-B_3 \frac{d+1}{2d^2}\frac{\Gamma \left(
\frac{d+1}{2} \right) }{\Gamma \left( \frac{d}{2}\right)}n^2\sigma^{d+1}\sqrt{mT}\chi (1+\alpha)\left(1-\frac{c_0}{32}\right)
\left[\frac{d}{d+2}\left(\partial_j U_i+\partial_i U_j-\frac{2}{d}\delta_{ij}
\nabla \cdot \mathbf{U}\right)+\delta_{ij}\nabla \cdot \mathbf{U}\right].
\end{eqnarray}
Note that in the last term of Eq.\ \eqref{a6} nonlinear terms in $c_0$ have been neglected. From Eq.\ \eqref{a6} one can easily identify the collisional transfer contributions to the hydrostatic pressure $p_c$, the shear viscosity $\eta_c$ and the bulk viscosity $\gamma$. They are given by
\begin{equation}
\label{a6.1}
p_c=2^{d-2}(1+\alpha)\phi \chi nT,
\end{equation}
\begin{equation}
\label{a6.2}
\gamma=\frac{2^{2(d-1)}\Gamma \left( \frac{d}{2}\right)}{\pi ^{\left( d+1\right) /2}}\phi^2\chi (1+\alpha)\left(1-\frac{c_0}{32}\right)\sigma^{1-d}\sqrt{mT},
\end{equation}
\begin{equation}
\label{a6.3}
\eta_c=\frac{2^{d-1}}{d+2}\phi \chi (1+\alpha)\eta_k+\frac{d}{d+2}\gamma,
\end{equation}
where use has been made of the definition \eqref{1.7} of the solid volume fraction $\phi$.


The collisional transfer contribution to the heat flux to first order in the gradients can be obtained in a similar way. The result is
\begin{eqnarray}
\label{a7}
q_i^c&=&\frac{1+\alpha}{2}m\sigma^{d}\chi \int\; \dd\mathbf{V}_1 \int\; \dd\mathbf{V}_2
\int \dd\widehat{\boldsymbol{\sigma}}\,\Theta (\widehat{{\boldsymbol {\sigma}}}
\cdot {\bf g})(\widehat{\boldsymbol {\sigma }}\cdot {\bf g})^2
(\widehat{\boldsymbol {\sigma }}\cdot {\bf G}) \widehat{\sigma}_i\left[f_1(\mathbf{V}_1)f_0(\mathbf{V}_2)+\frac{1}{2}f_0(\mathbf{V}_1)
\widehat{{\boldsymbol {\sigma}}}\cdot \nabla f_0(\mathbf{V}_2)\right]\nonumber\\
&=&\frac{3}{2}\frac{B_2}{d+2}n \sigma^d\chi q_i^k-\partial_i T\frac{B_3}{8d}\frac{m\sigma^{d+1}}{T}\chi (1+\alpha)\int\; \dd\mathbf{V}_1 \int\; \dd\mathbf{V}_2 \; f_0(\mathbf{V}_1)
f_0(\mathbf{V}_2)\left[g^{-1}({\bf g}\cdot {\bf G})^2+gG^2+\frac{3}{2}g({\bf g}\cdot {\bf G})+\frac{1}{4}g^3\right]\nonumber\\
&=&\frac{3}{2}\frac{B_2}{d+2}n \sigma^d\chi q_i^k-\partial_i T\frac{B_3}{2d}
\frac{\Gamma\left(\frac{d+3}{2}\right)}{\Gamma\left(\frac{d}{2}\right)}
\left(1+\frac{7}{32}c_0\right)n^2\sigma^{d+1}\sqrt{\frac{T}{m}}\chi (1+\alpha).
\end{eqnarray}
From Eq.\ \eqref{a7} one may identify the collisional contributions $\kappa_c$ and $\mu_c$ to the thermal conductivity $\kappa$ and the coefficient $\mu$, respectively. They are given by
\begin{equation}
\label{a7.1}
\kappa_c=3\frac{2^{d-2}}{d+2}\phi \chi (1+\alpha)\kappa_k+\frac{2^{2d-3}\Gamma \left( \frac{d}{2}\right)}{\pi ^{\left(d+1\right) /2}}\phi^2\chi (1+\alpha)\left(1+\frac{7}{32}c_0\right)\sigma^{1-d}\sqrt{\frac{T}{m}},
\end{equation}
\begin{equation}
\label{a7.2}
\mu_c=3\frac{2^{d-2}}{d+2}\phi \chi (1+\alpha)\mu_k.
\end{equation}

\section{Kinetic contributions to the fluxes}
\label{appB}

The kinetic contributions to the momentum and heat fluxes are defined by Eq.\ \eqref{2.10}. To obtain them in the NS approximation, one has first to expand the collision operator $J_\text{E}[f,f]$ to first order in the gradients. To do so the following results are needed:
\begin{equation}
\label{b1}
\chi({\bf r},{\bf r}\pm \boldsymbol{\sigma}|n)\to \chi \left( 1 \pm  \frac{1}{2}n\frac{\partial \ln \chi}{\partial n}
\boldsymbol{\sigma} \cdot \nabla \ln n \right),
\end{equation}
\begin{equation}
\label{b2}
f({\bf r}\pm \boldsymbol{\sigma}, \mathbf{V},t)\to f_0({\bf r}, \mathbf{V},t)+f_1({\bf r}, \mathbf{V},t)\pm \boldsymbol{\sigma}\cdot \nabla f_0({\bf r}, \mathbf{V},t),
\end{equation}
where $f_0$ and $f_1$ are defined in Eqs.\ \eqref{a2} and \eqref{a3}, respectively, and $\chi$ is obtained from the functional $\chi({\bf r},{\bf r}\pm \boldsymbol{\sigma}|n)$ by
evaluating all density fields at $n(\mathbf{r} ,t)$. The collision operator to first order then becomes\cite{GD99}
\begin{eqnarray}
\label{b3}
J_\text{E}[f,f]&\to& -\left(1+\frac{1}{2}\phi \frac{\partial \ln \chi}{\partial \phi}\right)
\boldsymbol{{\cal K}}[f_0] \cdot \nabla \ln n+\frac{1}{2}\boldsymbol{{\cal K}}\left[\frac{\partial}{\partial \mathbf{V}}\cdot (\mathbf{V}f_0)\right]\cdot \nabla  \ln T\nonumber\\
& & +\frac{1}{2} {\cal K}_i\left[\frac{\partial f_0}{\partial V_j}\right]\Delta_{ij}+\frac{1}{d} {\cal K}_i\left[\frac{\partial f_0}{\partial V_i}\right]\nabla \cdot \mathbf{U}
-\mathcal{L}f_1,
\end{eqnarray}
where
\begin{equation}
\label{b3.1}
\Delta_{ij}\equiv \partial_j U_i+\partial_i U_j-\frac{2}{d}\delta_{ij}\nabla \cdot \mathbf{U},
\end{equation}
and we have introduced the operators
\begin{equation}
\label{b4}
\mathcal{L}X=-\left(J^{(0)}[f_0,X]+J^{(0)}[X,f_0]\right),
\end{equation}
\begin{equation}
\label{b5}
J^{(0)}\left[X,Y\right]=\chi \sigma^{d-1}\int
\dd\mathbf{v}_{2}\int \dd\widehat{\boldsymbol {\sigma }}\Theta
(\widehat{\boldsymbol {\sigma}}\cdot \mathbf{g})(\widehat{
\boldsymbol {\sigma }}\cdot \mathbf{g})\left[ \alpha^{-2}X(\mathbf{v}_{1}^{\prime
})Y(\mathbf{v}_{2}^{\prime})-X(\mathbf{v}_{1})Y(\mathbf{v}_{2})\right],
\end{equation}
and
\begin{equation}
\boldsymbol{\mathcal{K}}[X] =\sigma^{d}\chi\int \dd \mathbf{v}_{2}\int \dd\widehat{\boldsymbol {\sigma
}}\Theta (\widehat{\boldsymbol {\sigma}} \cdot
\mathbf{g})(\widehat{\boldsymbol {\sigma }}\cdot
\mathbf{g})
\boldsymbol{\widehat{\sigma}}\left[ \alpha
^{-2}f_0(\mathbf{v} _{1}')X(\mathbf{v}_{2}')+f_0(\mathbf{v}_{1})X(\mathbf{v}_{2})\right].  \label{b6}
\end{equation}
Here, ${\bf v}_1^{\prime}$ and ${\bf v}_2'$ are defined by Eq.\ \eqref{2.3.1}.

We consider the kinetic contributions to the shear viscosity $\eta$. Multiply both sides of Eq.\ \eqref{2.1} by $mV_iV_j$ and integrate over velocity to get
\begin{equation}
\label{b7}
\partial_t P_{ij}^k+P_{ij}^k \nabla \cdot \mathbf{U}+U_\ell \partial_\ell P_{ij}^k+P_{\ell j}^k\partial_\ell U_i +
P_{\ell i}^k\partial_\ell U_j +\frac{2}{d+2}\partial_\ell(q_i^k\delta_{j\ell}+q_j^k\delta_{i\ell}+q_\ell^k\delta_{ij})=\int\; \dd\mathbf{v} m V_i V_j J_\text{E}[f,f],
\end{equation}
where use has been made of the result
\begin{equation}
\label{b8}
\int\; \dd\mathbf{v}\; m V_i V_j V_\ell f(\mathbf{V})= \frac{2}{d+2}(q_i^k\delta_{j\ell}+q_j^k\delta_{i\ell}+q_\ell^k\delta_{ij}).
\end{equation}
The integral on the right hand side of Eq.\ \eqref{b7} can performed by using the definition \eqref{b3}
\begin{equation}
\label{b9}
\int\; \dd\mathbf{v}\; m V_i V_j  J_\text{E}[f,f]=- \int\; \dd\mathbf{v}\; m V_i V_j \mathcal{L}f_1+
\frac{1}{2}\int\; \dd\mathbf{v}\; m V_i V_j  {\cal K}_\ell\left[\frac{\partial f_0}{\partial V_p}\right]
\Delta_{\ell k}
+\frac{1}{d}\int\; \dd\mathbf{v}\; m V_i V_j  {\cal K}_\ell\left[\frac{\partial f_0}{\partial V_\ell}\right]\nabla \cdot \mathbf{U}.
\end{equation}
The collision integrals appearing in Eq.\ \eqref{b9} can be evaluated by considering the explicit forms of $f_0$ and $f_1$. The first integral is given by \cite{GD99,RGD12,G12}
\begin{equation}
\label{b10}
\int\; \dd\mathbf{v}\; m V_i V_j \mathcal{L}f_1=\nu_\eta \Pi_{ij}+nT \zeta_0  \delta_{ij},
\end{equation}
where $\nu_\eta$ and $\zeta_0$ are given by Eqs.\ \eqref{4.8} and \eqref{4.9}, respectively. Note that nonlinear terms in $c_0$, $\Pi_{ij}$ and $\mathbf{q}^k$ have been neglected in Eq.\  \eqref{b10}  when $f_0$ and $f_1$ are replaced by their Grad's approximations \eqref{a2} and \eqref{a3}, respectively. The remaining two integrals can be performed using the definition of $\boldsymbol{{\cal K}}$ in Eq.\ \eqref{b6}. The first integral is
\begin{eqnarray}
\label{b11}
\int\; \dd\mathbf{v}\; m V_i V_j  {\cal K}_\ell\left[\frac{\partial f_0}{\partial V_p}\right]
\Delta_{\ell p}&=&\sigma^{d}\chi\int \dd \mathbf{V}_{1}\int \dd \mathbf{V}_{2}\;mV_{1i}V_{1j}\;\int \dd\widehat{\boldsymbol {\sigma
}}\Theta (\widehat{\boldsymbol {\sigma}} \cdot
\mathbf{g})(\widehat{\boldsymbol {\sigma }}\cdot
\mathbf{g})\widehat{\sigma}_\ell\nonumber\\
& &\times\left[ \alpha
^{-2}f_0(\mathbf{V} _{1}')\frac{\partial f_0(\mathbf{V}_2')}{\partial V_{2p}'}
+f_0(\mathbf{V}_{1})\frac{\partial f_0(\mathbf{V}_2)}{\partial V_{2p}}\right]\Delta_{\ell p}.
\end{eqnarray}
A simpler form of this integral is obtained by changing variables to integrate over $\mathbf{V}_1'$ and $\mathbf{V}_2'$ instead of $\mathbf{V}_1$ and $\mathbf{V}_2$ in the first term of Eq.\ \eqref{b11}. The Jacobian of the transformation is $\alpha$ and $\widehat{\boldsymbol {\sigma}} \cdot\mathbf{g}=-\alpha \widehat{\boldsymbol {\sigma}} \cdot\mathbf{g}'$. Also, $\mathbf{V}_1(\mathbf{V}_1',\mathbf{V}_2')\equiv \mathbf{V}_1''=\mathbf{V}_1-\frac{1}{2}(1+\alpha)\widehat{\boldsymbol {\sigma}}(\widehat{\boldsymbol {\sigma}} \cdot\mathbf{g})$. The integral then becomes
\begin{eqnarray}
\label{b11.1}
\int \dd\mathbf{v} m V_i V_j  {\cal K}_\ell\left[\frac{\partial f_0}{\partial V_p}\right]
\Delta_{\ell p}&=&-m\chi \sigma^d\int \dd \mathbf{V}_{1}\int \dd \mathbf{V}_{2}\;f_0(\mathbf{V}_1)
\frac{\partial f_0(\mathbf{V}_2)}{\partial V_{2p}}\int\;\dd\widehat{\boldsymbol {\sigma
}}\Theta (\widehat{\boldsymbol {\sigma}} \cdot
\mathbf{g})(\widehat{\boldsymbol {\sigma }}\cdot
\mathbf{g})\widehat{\sigma}_\ell\left(V_{1i}''V_{1j}''-V_{1i}V_{1j}\right)\Delta_{\ell p}\nonumber\\
&=&
m\chi\frac{1+\alpha}{4} \sigma^d \int \dd \mathbf{V}_{1}\int \dd \mathbf{V}_{2}\;f_0(\mathbf{V}_1)
\frac{\partial f_0(\mathbf{V}_2)}{\partial V_{2p}}\int\;\dd\widehat{\boldsymbol {\sigma
}}\Theta (\widehat{\boldsymbol {\sigma}} \cdot
\mathbf{g})(\widehat{\boldsymbol {\sigma }}\cdot
\mathbf{g})^2\nonumber\\
& & \times\left[2\left(V_{1i}\widehat{\sigma}_j\widehat{\sigma}_\ell+V_{1j}\widehat{\sigma}_i
\widehat{\sigma}_\ell\right)-(1+\alpha)(\widehat{\boldsymbol {\sigma}} \cdot
\mathbf{g})\widehat{\sigma}_i\widehat{\sigma}_j\widehat{\sigma}_\ell\right]\Delta_{\ell p}
\nonumber\\
&=&m\chi\frac{1+\alpha}{4} \sigma^d\;\int \dd \mathbf{V}_{1}\int \dd \mathbf{V}_{2}\;f_0(\mathbf{V}_1)
f_0(\mathbf{V}_2)\int\;\dd\widehat{\boldsymbol {\sigma
}}\Theta (\widehat{\boldsymbol {\sigma}} \cdot
\mathbf{g})(\widehat{\boldsymbol {\sigma }}\cdot\mathbf{g})\nonumber\\
& & \times\left[4\left(V_{1i}\widehat{\sigma}_j\widehat{\sigma}_\ell\widehat{\sigma}_p
+V_{1j}\widehat{\sigma}_i\widehat{\sigma}_\ell\widehat{\sigma}_p\right)-3(1+\alpha)(\widehat{\boldsymbol {\sigma}} \cdot\mathbf{g})\widehat{\sigma}_i\widehat{\sigma}_j\widehat{\sigma}_\ell\widehat{\sigma}_p\right]
\Delta_{\ell p}\nonumber\\
&=&\frac{2^{d-1}}{d+2}nT\phi \chi (1+\alpha)(1-3\alpha)\Delta_{ij},
\end{eqnarray}
where in the last step use has been made of the angular integrals
\begin{equation}
\label{b11.2}
\int\;\dd\widehat{\boldsymbol {\sigma
}}\Theta (\widehat{\boldsymbol {\sigma}} \cdot
\mathbf{g})(\widehat{\boldsymbol {\sigma }}\cdot\mathbf{g})\widehat{\sigma}_i\widehat{\sigma}_j\widehat{\sigma}_\ell=\frac{B_2}{d+2}\left(
\delta_{ij}g_\ell+\delta_{i\ell}g_j+\delta_{j\ell}g_i \right),
\end{equation}
\begin{eqnarray}
\label{b11.3}
\int\;\dd\widehat{\boldsymbol {\sigma
}}\Theta (\widehat{\boldsymbol {\sigma}} \cdot
\mathbf{g})(\widehat{\boldsymbol {\sigma }}\cdot\mathbf{g})^2\widehat{\sigma}_i\widehat{\sigma}_j\widehat{\sigma}_\ell\widehat{\sigma}_p&=&\frac{B_2}
{(d+2)(d+4)}\left[2\left(g_ig_j\delta_{\ell p}+g_ig_\ell\delta_{pj}+g_ig_p\delta_{\ell j}+
g_\ell g_j\delta_{i p}+g_pg_j\delta_{i \ell}+g_\ell g_p\delta_{ij}\right)\right.\nonumber\\
& & \left. +g^2\left(\delta_{ij}\delta_{\ell p}+\delta_{i\ell}\delta_{jp}+\delta_{ip}\delta_{j\ell}\right)\right].
\end{eqnarray}
The last integral appearing in Eq.\ \eqref{b9} can be performed by using similar mathematical steps. It is given by
\begin{equation}
\label{b12}
\int\; \dd{\bf v}\;m V_iV_j \mathcal{K}_\ell\left[\frac{\partial f_0}{\partial V_\ell}
\right]=\delta_{ij}2^{d-2}nT\chi \phi (1+\alpha)(1-3\alpha).
\end{equation}

We are interested in the solution to Eq.\ \eqref{b7} in the NS approximation. Thus, in order to solve \eqref{b7}, one needs to make use of the balance equations \eqref{2.7}--\eqref{2.9} up to first order in the spatial gradients:
\begin{equation}
\label{b13}
\partial_t n \to -\mathbf{U}\cdot \nabla n-n \nabla \cdot \mathbf{U}, \quad
\partial_t \mathbf{U} \to - \mathbf{U}\cdot \nabla \mathbf{U}-\rho^{-1}\nabla p,\quad
\partial_t T \to -\mathbf{U}\cdot \nabla T-\frac{2}{dn}p \nabla \cdot \mathbf{U}-\zeta T,
\end{equation}
where the pressure $p$ is given by Eq.\ \eqref{4.1} and the cooling rate can be written as (see Appendix \ref{appC})
\begin{equation}
\label{b14}
\zeta\to \zeta_0+\left(\zeta_{10}+\zeta_{11}  c_1\right)\nabla \cdot \mathbf{U}.
\end{equation}
According to Eq.\ \eqref{1.2}, the kinetic contribution $\eta_k$ to the shear viscosity is defined as
\begin{equation}
\label{b15}
\Pi_{ij}=-\eta_k \Delta_{ij}.
\end{equation}
The coefficient $\eta_k$ can be easily obtained from Eq.\ \eqref{b7} when one takes into account Eqs.\ \eqref{b9}, \eqref{b10}, \eqref{b11.1}, and \eqref{b12}. The corresponding equation for $\eta_k$ is
\begin{equation}
\label{b16}
\left(\partial_t+\nu_\eta\right) \eta_k=n T\left[1-\frac{2^{d-2}}{d+2}\chi \phi (1+\alpha)(1-3\alpha)\right],
\end{equation}
where the time derivative $\partial_t \eta_k$ must be evaluated at zeroth-order in spatial gradients. From dimensional analysis $\eta_k \propto T^{1/2}$ and so,
\begin{equation}
\label{b16.1}
\partial_t \eta_k=\frac{1}{2}\eta_k \partial_t \ln T=-\frac{1}{2}\zeta_0 \eta_k.
\end{equation}
The solution to Eq.\ \eqref{b16} is given by Eq.\ \eqref{4.7} when one takes into account the result \eqref{b16.1}.

Apart from obtaining $\eta_k$, the coefficients in Eq.\ \eqref{b7} proportional to the divergence of the flow velocity allows one to determine $\zeta_{10}$. It is given by
\begin{equation}
\label{b17}
\zeta_{10}=-3\frac{2^{d-2}}{d}\chi \phi (1-\alpha^2).
\end{equation}
This first-order contribution to the cooling rate will be also determined in the Appendix \ref{appC} by following a different route. Moreover, the coefficient $c_1=0$ at this level of approximation. A nonzero contribution to $c_1$ will be obtained when we determine the (contracted) fourth degree velocity moment of $f$.

The kinetic parts of the thermal conductivity $\kappa$ and the coefficient $\mu$ are obtained in a similar way. Multiplication of Eq.\ \eqref{2.1} by $\frac{m}{2}V^2V_i$ and integration over the velocity yields
\begin{equation}
\label{b18}
\partial_t q_i^k+\frac{d+2}{2}\left[nT\partial_t U_i+
\frac{c_0}{2m}\partial_i(n T^2)\right]+\frac{d+2}{2m}\partial_j\left(n T^2\right)+\frac{d+2}{2}nT{\bf U}\cdot \nabla U_i=
\int\; \dd\mathbf{v} \frac{m}{2}V^2V_i J_\text{E}[f,f],
\end{equation}
where only linear terms in the spatial gradients have been considered on the left hand side of Eq.\ \eqref{b18}. In addition, I have assumed that $c_0$ is uniform (see Eq.\ \eqref{4.20}) and have used the relation
\begin{equation}
\label{b18.1}
\int\; \dd\mathbf{v} \frac{m}{2}V^2V_i f(\mathbf{V})=\frac{d+2}{4}\frac{nT^2}{m}\delta_{ij}
\left(c_0+c_1 \nu^{-1}\nabla \cdot \mathbf{U}\right)+\frac{T}{m}\left(\frac{d+4}{2}P_{ij}^k-
nT\delta_{ij}\right).
\end{equation}
Equation \eqref{b18} can be more explicitly written when one takes into account the balance equations \eqref{b13}. The result is
\begin{equation}
\label{b18.2}
\partial_t q_i^k+\frac{d+2}{2}\frac{T^2}{m}\left(1+\frac{c_0}{2}-p^*-\phi \frac{\partial p^*}{\partial \phi}\right)\partial_i n+\frac{d+2}{2}\frac{nT}{m}\left(2+c_0-p^*\right)\partial_i T=\int\; \dd\mathbf{v} \frac{m}{2}V^2V_i J_\text{E}[f,f],
\end{equation}
where
\begin{equation}
\label{b18.3}
p^*\equiv \frac{p}{nT}=1+2^{d-2}\phi \chi (1+\alpha).
\end{equation}
The collision integral on the right side of \eqref{b18.2} can be evaluated by using similar mathematical steps as those made before in Eqs.\ \eqref{b10}--\eqref{b12}. After a tedious and long algebra, one gets \cite{GD99,RGD12,G12}
\begin{eqnarray}
\label{b19}
\int\; \dd\mathbf{v} \frac{m}{2}V^2V_i J_\text{E}[f,f]&=&-\int\; \dd\mathbf{v} \frac{m}{2}V^2V_i
\mathcal{L}f_1
-\left(1+\frac{1}{2}\phi \frac{\partial \ln \chi}{\partial \phi}\right)\partial_j \ln n\;\int\; \dd\mathbf{v} \frac{m}{2}\;V^2\;V_i\;
{\cal K}_j[f_0] \nonumber\\
& & +\frac{1}{2}\partial_j \ln T\;\int\; \dd\mathbf{v} \frac{m}{2}\;V^2\;V_i\;{\cal K}_j\left[\frac{\partial}{\partial \mathbf{V}}\cdot (\mathbf{V}f_0
)\right] \nonumber\\
&=&-\nu_\kappa q_i^k-2^{d-3}\phi \chi (1+\alpha)\frac{nT^2}{m}\left(1+\frac{1}{2}\phi \frac{\partial \ln \chi}{\partial \phi}\right)\nonumber\\
& & \times
\left[2(d+2)+3\alpha(\alpha-1)+\frac{c_0}{4}(10+2d-3\alpha+3\alpha^2)\right]
\partial_i \ln n \nonumber\\
& & -2^{d-4}\phi \chi (1+\alpha)\frac{nT^2}{m}\left[2(d+2)+3(1+\alpha)(2\alpha-1)+\frac{3}{2}c_0(1+\alpha)^2\right]
\partial_i \ln T,
\end{eqnarray}
where $\nu_\kappa$ is given by Eq.\ \eqref{4.15}. As before, linear terms in $c_0$, $\Pi_{ij}$, $\mathbf{q}^k$ and $c_1$ have been only retained in Eq.\ \eqref{b19}.

The constitutive equation for $\mathbf{q}^k$ is
\begin{equation}
\label{b20}
\mathbf{q}^k=-\kappa_k \nabla T-\mu_k \nabla n.
\end{equation}
Dimensional analysis shows that $\kappa_k \propto T^{1/2}$ and $\mu_k \propto T^{3/2}$. Thus, according to Eq.\ \eqref{b20},  to first order in the gradients the time derivative of the kinetic contribution to the heat flux can be written as
\begin{equation}
\label{b21}
\partial_t q_i^k=2\zeta_0 \kappa_k \partial_i T+\zeta_0\left[\frac{T\kappa_k}{n}\left(1+\phi \frac{\partial \ln \chi}{\partial \phi}\right)+\frac{3}{2}\mu_k \right] \partial_i n,
\end{equation}
where use has been made of the relation
\begin{equation}
\label{b22}
\partial_i(\partial_t T)\to -\partial_i(\zeta_0 T)=-\zeta_0 T \left(1+\phi \frac{\partial \ln \chi}{\partial \phi}\right)\partial_i \ln n-\frac{3}{2}\zeta_0 T\partial_i \ln T.
\end{equation}
Substitution of Eqs.\ \eqref{b19} and \eqref{b21} into Eq.\ \eqref{b18.2} leads to the following set of equations when one equates coefficients pertaining to the density and temperature gradients:
\begin{equation}
\label{b23}
2\zeta_0 \kappa_k+\frac{d+2}{2}\frac{nT}{m}\left(2+c_0-p^*\right)=\nu_\kappa \kappa_k-
2^{d-4}\phi \chi (1+\alpha)\frac{nT}{m}\left[2(d+2)+3(1+\alpha)(2\alpha-1)+\frac{3}{2}c_0(1+\alpha)^2\right],
\end{equation}
\begin{eqnarray}
\label{b24}
& & \zeta_0 \left[ \frac{T\kappa_k}{n}\left(1+\phi \frac{\partial \ln \chi}{\partial \phi}\right)
+\frac{3}{2}\mu_k\right]+\frac{d+2}{2}\frac{T^2}{m}\left(1+\frac{c_0}{2}-p^*-\phi\frac{\partial p^*}{\partial \phi}\right)=\nu_\kappa \mu_k\nonumber\\
& & -
2^{d-3}\phi \chi (1+\alpha)\frac{T^2}{m}\left(1+\frac{1}{2}\phi \frac{\partial \ln \chi}{\partial \phi}\right)\left[2(d+2)+3\alpha(\alpha-1)+\frac{c_0}{4}(10+2d-3\alpha+3\alpha^2)\right].
\end{eqnarray}

The solution to Eqs.\ \eqref{b23} and \eqref{b24} gives the expressions \eqref{4.12} and \eqref{4.13} for $\kappa_k$ and $\mu_k$, respectively, once the explicit form \eqref{b18.3} for $p^*$ is considered.

\section{Cooling rate $\zeta$ and fourth moment $c$}
\label{appC}

In this Appendix, the cooling rate $\zeta$ and the fourth moment $c$ are determined to first order in gradients. The cooling rate $\zeta$ is defined by Eq.\ \eqref{2.13}. Up to the first order in gradients, $\zeta$ is
\begin{eqnarray}
\label{c1}
\zeta&=&\frac{1-\alpha^2}{4dnT}m\sigma^{d-1}\chi \int\; \dd\mathbf{V}_1 \int\; \dd\mathbf{V}_2
\int \dd\widehat{\boldsymbol{\sigma}}\,\Theta (\widehat{{\boldsymbol {\sigma}}}
\cdot {\bf g})(\widehat{\boldsymbol {\sigma }}\cdot {\bf g})^3
f_0(\mathbf{V}_1)f_0(\mathbf{V}_2)\nonumber\\
& &-\frac{1-\alpha^2}{4dnT}m\sigma^{d}\chi \int\; \dd\mathbf{V}_1 \int\; \dd\mathbf{V}_2\; f_0(\mathbf{V}_1)f_0(\mathbf{V}_2)
\int \dd\widehat{\boldsymbol{\sigma}}\,\Theta (\widehat{{\boldsymbol {\sigma}}}
\cdot {\bf g})(\widehat{\boldsymbol {\sigma }}\cdot {\bf g})^3
\left(\frac{\partial \ln f_0(\mathbf{V}_2)}{\partial V_{2i}}\right)(\widehat{\boldsymbol {\sigma }}\cdot \nabla)U_i\nonumber\\
& & +\frac{1-\alpha^2}{4dnT}m\sigma^{d-1}\chi \int\; \dd\mathbf{V}_1 \int\; \dd\mathbf{V}_2\left[
f_0(\mathbf{V}_1)f_1(\mathbf{V}_2)+f_1(\mathbf{V}_1)f_0(\mathbf{V}_2)\right]\int \dd\widehat{\boldsymbol{\sigma}}\,\Theta (\widehat{{\boldsymbol {\sigma}}}
\cdot {\bf g})(\widehat{\boldsymbol {\sigma }}\cdot {\bf g})^3
\nonumber\\
&=&\zeta_0+\zeta_{10}\nabla \cdot \mathbf{U}+\zeta_{11} c_1 \nabla \cdot \mathbf{U},
\end{eqnarray}
where in the last step use has been made of the form \eqref{3.5} of the fourth moment $c$. In Eq.\ \eqref{c1},
\begin{equation}
\label{c2}
\zeta_0\equiv B_3 \frac{1-\alpha^2}{4dnT}m\sigma^{d-1}\chi \int\; \dd\mathbf{V}_1 \int\; \dd\mathbf{V}_2\;
g^3\;f_0(\mathbf{V}_1)f_0(\mathbf{V}_2),
\end{equation}
\begin{equation}
\label{c3}
\zeta_{10}\equiv -\frac{3B_2}{d} \frac{1-\alpha^2}{4dnT}m\sigma^{d-1}\chi \int\; \dd\mathbf{V}_1 \int\; \dd\mathbf{V}_2\;
g^2\;f_0(\mathbf{V}_1)f_0(\mathbf{V}_2),
\end{equation}
\begin{equation}
\label{c4}
\zeta_{11}\equiv B_3 \frac{1-\alpha^2}{8dnT}m\sigma^{d-1}\chi \nu^{-1} \int\; \dd\mathbf{V}_1 \int\; \dd\mathbf{V}_2\;
g^3\; E(\mathbf{V}_2)f_0(\mathbf{V}_1)f_0(\mathbf{V}_2).
\end{equation}
Here, the coefficients $B_n$ are defined by Eq.\ \eqref{a5} and $\nu$ is given by Eq.\ \eqref{4.8.1}. The integration over velocity can be carried out in Eqs.\ \eqref{c2}--\eqref{c4} when one takes into account the expression \eqref{a2} of $f_0$. Neglecting nonlinear terms in $c_0$, the coefficients $\zeta_0$ and $\zeta_{10}$ are given by Eqs.\ \eqref{4.9} and \eqref{b17}, respectively while $\zeta_{11}$ is
\begin{equation}
\label{c5}
\zeta_{11}=\frac{3(d+2)}{128d}(1-\alpha^2)\chi\left(1+\frac{3c_0}{64}\right).
\end{equation}

The complete determination of the cooling rate and the NS transport coefficients still requires to get the quantities $c_0$ and $c_1$. To evaluate them, one multiplies both sides of Eq.\ \eqref{2.1} by $V^4$ and integrates over velocity. In the NS order, one gets
\begin{equation}
\label{c5.1}
\frac{d(d+2)}{m^2}\left(1+\frac{c_0}{2}+\frac{c_1}{2\nu}\nabla \cdot \mathbf{U} \right)\left[\partial_t(nT^2)+
\frac{d+4}{d}nT^2\nabla\cdot \mathbf{U}+\mathbf{U}\cdot \nabla (nT^2)\right]=\int\;\dd\mathbf{v} V^4 J_\text{E}[f,f].
\end{equation}
The left hand side of Eq.\ \eqref{c5.1} can be simplified when one takes into account the balance equations \eqref{b13}. To first order in spatial gradients, the result is
\begin{equation}
\label{c6}
-2d(d+2)\frac{nT^2}{m^2}\left\{\left(1+\frac{c_0}{2}\right)\left[\zeta_0+\left(\frac{2}{d}(p^*-1)+\zeta_{10}+
\zeta_{11}c_1\right)\nabla \cdot \mathbf{U}\right]+\frac{\zeta_0}{2\nu}c_1\nabla \cdot \mathbf{U}\right\}=
\int\;\dd\mathbf{v} V^4 J_\text{E}[f,f].
\end{equation}
The collision integral on the right side can be computed by using similar mathematical steps as those made before. Neglecting nonlinear terms in $c_0$ and $c_1$ and after a tedious algebra, one gets
\begin{eqnarray}
\label{c7}
\int\;\dd\mathbf{v} V^4 J_\text{E}[f,f]&=&-\frac{d+2}{2}\frac{nT^2}{m^2} (1-\alpha^2)\left[d+\frac{3}{2}+\alpha^2
+\frac{c_0}{2}\left(\frac{3}{32}(10d+39+10\alpha^2)+\frac{d-1}{1-\alpha}\right)\right]\nu \nonumber\\
& &+\left(\frac{c_1}{4}\nu_\zeta+ \lambda  \right)\frac{nT^2}{m^2}\nabla \cdot \mathbf{U},
\end{eqnarray}
where $\lambda$ and $\nu_\zeta$ are given by Eqs.\ \eqref{4.18} and \eqref{4.19}, respectively. Upon deriving Eq.\ \eqref{c7}, use has been made of the partial results
\begin{equation}
\label{c7.1}
\int\; \dd\mathbf{v}\;V^4 {\cal L}f_1=\nu_\zeta \frac{nT^2}{m^2},
\end{equation}
\begin{equation}
\label{c7.2}
\int\; \dd{\bf v}\;V^4 \mathcal{K}_\ell\left[\frac{\partial f_0}{\partial V_\ell}
\right]=d \lambda \frac{nT^2}{m^2}.
\end{equation}
Substitution of Eq.\ \eqref{c7} into Eq.\ \eqref{c6} allows one to explicitly determine $c_0$ and $c_1$. They are given by Eqs.\ \eqref{4.20} and \eqref{4.17}, respectively. Note that in Eqs.\ \eqref{4.20} and \eqref{4.17} nonlinear terms in $c_0$ and $c_1$ (like $c_0^2$, $c_0c_1$ and $c_1^2$) have been neglected.

\section{Jenkins and Richman expressions}
\label{appD}

The expressions obtained years ago by Jenkins and Richman \cite{JR85a,JR85b} for the NS transport coefficients of a dense gas of inelastic hard diks \cite{JR85a} and spheres \cite{JR85b} are displayed in this Appendix. These authors solved the Enskog kinetic equation from thirteen Grad's moment method.

In the case of smooth hard disks ($d=2$), their results are \cite{JR85a}
\begin{equation}
\label{d1}
\eta=\frac{8\eta_0}{\chi(7-3\alpha)(1+\alpha)}\left[1-\frac{1}{4}(1+\alpha)(1-3\alpha)\phi \chi \right]
\left[1+\frac{1}{2}(1+\alpha)\phi \chi \right]+\frac{1}{2}\gamma,
\end{equation}
\begin{equation}
\label{d2}
\gamma=\frac{8}{\pi}\phi^2\chi (1+\alpha)\eta_0,
\end{equation}
\begin{equation}
\label{d3}
\kappa=\frac{2\kappa_0}{\chi(1+\alpha)\left[1+\frac{15}{4}(1-\alpha)\right]}
\left[1+\frac{3}{8}(1+\alpha)^2(2\alpha-1)\phi \chi\right]
\left[1+\frac{3}{4}(1+\alpha)\phi \chi\right]+
\frac{2}{\pi}\phi^2\chi (1+\alpha)\kappa_0,
\end{equation}
\begin{equation}
\label{d3}
\mu=-\frac{3T\kappa_0}{2n\chi(1+\alpha)\left[1+\frac{15}{4}(1-\alpha)\right]}
\phi \chi \left(1+\frac{1}{2}\phi \frac{\partial \ln \chi}{\partial \phi}\right)
\alpha(1-\alpha^2)\left[1+\frac{3}{4}\phi \chi (1+\alpha)\right].
\end{equation}
Here, $\eta_0$ and $\kappa_0$ are given by Eqs.\ \eqref{4.6} and \eqref{4.14}, respectively.

In the case of smooth hard spheres ($d=3$), their results are \cite{JR85b}
\begin{equation}
\label{d4}
\eta=\frac{4\eta_0}{\chi(3-\alpha)(1+\alpha)}\left[1-\frac{2}{5}(1+\alpha)(1-3\alpha)\phi \chi \right]
\left[1+\frac{4}{5}(1+\alpha)\phi \chi \right]+\frac{3}{5}\gamma,
\end{equation}
\begin{equation}
\label{d5}
\gamma=\frac{128}{5\pi}\phi^2\chi (1+\alpha)\eta_0,
\end{equation}
\begin{equation}
\label{d6}
\kappa=\frac{2\kappa_0}{\chi(1+\alpha)\left[1+\frac{33}{16}(1-\alpha)\right]}
\left[1+\frac{3}{5}(1+\alpha)^2(2\alpha-1)\phi \chi\right]
\left[1+\frac{6}{5}(1+\alpha)\phi \chi\right]+
\frac{256}{25\pi}\phi^2\chi (1+\alpha)\kappa_0,
\end{equation}
\begin{equation}
\label{d7}
\mu=-\frac{12T\kappa_0}{5n\chi(1+\alpha)\left[1+\frac{33}{16}(1-\alpha)\right]}
\phi \chi \left(1+\frac{1}{2}\phi \frac{\partial \ln \chi}{\partial \phi}\right)
\alpha(1-\alpha^2)\left[1+\frac{6}{5}\phi \chi (1+\alpha)\right].
\end{equation}


\begin{thebibliography} {99}
\bibitem{note}The expressions of the NS transport coefficients posted in Ref.\ \onlinecite{G12} are just a refinement of those derived by Lutsko in Ref.\ \onlinecite{L05}. Both theories are expected to differ for very small values of the coefficient of restitution.
\end{thebibliography}
\end{document}